\documentclass[11pt]{iopart}
\usepackage{iopams}  
\usepackage{revsymb}
\usepackage{graphicx}
\begin{document}

\newcommand{\apj}{{Astrophys.\ J. }}
\newcommand{\apjs}{{Astrophys.\ J.\ Supp. }}
\newcommand{\apjl}{{Astrophys.\ J.\ Lett. }}
\newcommand{\aj}{{Astron.\ J. }}
\newcommand{\prl}{{Phys.\ Rev.\ Lett. }}
\newcommand{\prd}{{Phys.\ Rev.\ D }}
\newcommand{\mnras}{{Mon.\ Not.\ R.\ Astron.\ Soc. }}
\newcommand{\araa}{{ARA\&A }}
\newcommand{\aap}{{Astron.\ \& Astrophy. }}
\newcommand{\nat}{{Nature }}
\newcommand{\cqg}{{Class.\ Quantum Grav.\ }}

\def\HS{{\mathfrak H}_3}
\def\is{\hbox{\scriptsize\rm i}}
\def\i{\hbox{\rm i}}
\def\j{\hbox{\rm j}}
\def\3{{\ss}}

\hfill astro-ph/0412569

\title[Poincar\'e Dodecahedron]
{CMB Anisotropy of the Poincar\'e Dodecahedron}

\author{Ralf Aurich, Sven Lustig, Frank Steiner}

\address{Abteilung Theoretische Physik, Universit\"at Ulm,\\
Albert-Einstein-Allee 11, D-89069 Ulm, Germany}

\begin{abstract}
We analyse the anisotropy of the cosmic microwave background (CMB)
for the Poincar\'e dodecahedron
which is an example for a multi-connected spherical universe.
We compare the temperature correlation function
and the angular power spectrum for the Poincar\'e dodecahedral universe
with the first-year WMAP data and find
that this multi-connected universe can explain the surprisingly
low CMB anisotropy on large scales found by WMAP
provided that the total energy density parameter
$\Omega_{\hbox{\scriptsize tot}}$ is in the range $1.016\dots 1.020$.
The ensemble average over the primordial perturbations is assumed
to be the scale-invariant Harrison-Zel'dovich spectrum.
The circles-in-the-sky signature is studied and it is found
that the signal of the six pairs of matched circles
could be missed by current analyses of CMB sky maps.
\end{abstract}

\pacs{98.80.-k, 98.70.Vc, 98.80.Es}




\section{Introduction}

The question of the global geometry of the Universe,
i.\,e.\ of its spatial curvature, topology and thus its shape,
is a very fundamental one.
While the problem of the spatial curvature is discussed in all
textbooks on cosmology,
mainly in connection with the total mass/energy densities
in the Universe and the various inflationary scenarios,
the possibility of a multi-connected Universe is only rarely mentioned.
This is surprising in view of the fact that it was already realized
before Einstein's seminal first paper on cosmology \cite{Einstein_1917}
that the three-dimensional space of astronomy might be not only
non-Euclidean, but also multi-connected
according to a given Clifford-Klein space form
(see e.\,g.\ \cite{Schwarzschild_1900}).
If the Universe is assumed to be simply connected,
the cosmological principle implies that it is not only
locally, but also globally isotropic and homogeneous.
This means that all spatial points are geometrically equivalent.
In the case of a multi-connected Universe, however,
the cosmological principle holds in general only on the universal
covering space of constant curvature,
i.\,e.\ on ${\cal E}^3$, ${\cal S}^3$ and ${\cal H}^3$ for
a flat, positively and negatively curved Universe, respectively.
While the Einstein gravitational field equations still hold
in a multi-connected Universe, the global structure of the Universe
at large scales will be more complicated.

The discovery of the temperature fluctuations $\delta T$
of the cosmic microwave background radiation (CMB) by COBE
in 1992 \cite{Smoot_et_al_1992} and the detailed measurements
by WMAP \cite{Bennett_et_al_2003} and by other groups
have led to a renewed interest in the question of the global geometry
of the Universe
(see \cite{Lachieze-Rey_Luminet_1995,Levin_2002} for reviews).

The most promising signatures to detect a possible
multi-connected spatial structure of the Universe are the
strange suppression of the CMB quadrupole and octopole,
and of the temperature correlation function
first observed by COBE \cite{Hinshaw_et_al_1996} and
nicely confirmed by WMAP \cite{Bennett_et_al_2003},
and the so-called circles-in-the-sky-signature proposed in
\cite{Cornish_Spergel_Starkman_1998b}.

Until about 1999, all observational data were consistent with the fact
that the total energy density $\varepsilon_{\hbox{\scriptsize tot}}$
(at the present epoch) amounted only to ca.\ 30\% of the critical energy
density $\varepsilon_{\hbox{\scriptsize crit}} = \frac{3 H_0 c^2}{8\pi G}$,
i.\,e.\ $\Omega_{\hbox{\scriptsize tot}} :=
\varepsilon_{\hbox{\scriptsize tot}}/\varepsilon_{\hbox{\scriptsize crit}}
\simeq 0.3 < 1$,
and thus strongly indicated that the geometry of the Universe is hyperbolic.
Consequently, several groups performed detailed studies of
compact and non-compact models of the Universe possessing
hyperbolic geometry.
Although later observations on the magnitude redshift relation of the
supernovae of type Ia \cite{Riess_et_al_1998,Perlmutter_et_al_1999}
provided evidence for a large dark energy component,
and the determination of the first acoustic peak in the CMB anisotropy
by TOCO \cite{Torbet_et_al_1999,Miller_et_al_1999},
BOOMERanG \cite{deBernardis_et_al_2000} and
MAXIMA-1 \cite{Hanany_et_al_2000,Balbi_et_al_2000}
indicated a nearly flat Universe,
i.\,e.\ $\Omega_{\hbox{\scriptsize tot}}\simeq 1$,
the available data are still compatible with the spatial geometry
of the Universe being hyperbolic.
(For a discussion of hyperbolic universes, see
\cite{Bond_Pogosyan_Souradeep_1998,Bond_Pogosyan_Souradeep_1999a,%
Bond_Pogosyan_Souradeep_1999b,Cornish_Spergel_Starkman_1998,Aurich_1999,%
Cornish_Spergel_1999,Inoue_Tomita_Sugiyama_1999,Aurich_Steiner_2000,%
Aurich_Steiner_2002b,Aurich_Steiner_2003,Aurich_Lustig_Steiner_Then_2004a,%
Aurich_Lustig_Steiner_Then_2004b}.)

The {\it concordance model} of cosmology assumes an exactly (spatially) flat
Universe with the topology of ${\cal E}^3$ and, furthermore,
that the dark energy is given by a positive cosmological constant $\Lambda$,
i.\,e.\ $\Omega_\Lambda := \frac{\Lambda c^2}{3H_0^2} =
1-\Omega_{\hbox{\scriptsize mat}}-\Omega_{\hbox{\scriptsize rad}}$
with $\Omega_{\hbox{\scriptsize mat}}=\Omega_{\hbox{\scriptsize bar}}+
\Omega_{\hbox{\scriptsize cdm}}$, where the various parameters denote
the present values of the baryonic (bar), cold dark matter (cdm),
matter (mat) and radiation (rad) energy densities in units of
$\varepsilon_{\hbox{\scriptsize crit}}$ ($\Lambda$CDM model).
In the following, we shall refer to three variants of the concordance
model presented by the WMAP team on the Legacy Archive for Microwave
Background Data Analysis (LAMBDA) Web site 
{\sl http://lambda.gsfc.nasa.gov}, see also \cite{Spergel_et_al_2003}.
These models give a good overall fit to the CMB anisotropy
on small and medium scales, but there remains a strange discrepancy
at large scales as first observed by COBE \cite{Hinshaw_et_al_1996} and
later substantiated by WMAP \cite{Bennett_et_al_2003}.

Let us expand the {\it temperature fluctuations} $\delta T(\hat n)$
of the microwave sky, where the dipole contribution has been subtracted,
into real spherical harmonics $\tilde Y_{lm}(\hat n)$
on ${\cal S}^2$,
\begin{equation}
\label{Eq:delta_T_expansion}
\delta T(\hat n) \; := \;
\sum_{l=2}^\infty \sum_{m=-l}^l \, a_{lm} \, \tilde Y_{lm}(\hat n)
\hspace{10pt} ,
\end{equation}
where $\hat n$ denotes the unit vector in the direction from
which the photons arrive.
(Note that the monopole and dipole terms, $l=0,1$, are not included
in the sum (\ref{Eq:delta_T_expansion}).)
From the real expansion coefficients $a_{lm}$ one forms
the {\it multipole moments}
\begin{equation}
\label{Eq:C_l}
C_l \; := \;
\frac 1{2l+1} \, \Big< \sum_{m=-l}^l \, \big( a_{lm} \big)^2 \, \Big>
\end{equation}
and the {\it angular power spectrum}
\begin{equation}
\label{Eq:angular_power_spectrum}
\delta T_l^2 \; := \; \frac{l(l+1)}{2\pi} \, C_l
\hspace{10pt} .
\end{equation}
The average $\left< \dots \right>$ in (\ref{Eq:C_l}) denotes an
ensemble average over the primordial perturbations to be discussed
in section \ref{angular_power_spectrum_and_correlation_function},
respectively an ensemble average over the universal observers.
The {\it temperature two-point correlation function} $C(\vartheta)$ is defined
as $C(\vartheta) := \left< \delta T(\hat n) \delta T(\hat n')\right>$
with $\hat n \cdot \hat n' = \cos\vartheta$,
which can be computed from the multipole moments (\ref{Eq:C_l})
under the assumption of statistical isotropy as
\begin{equation}
\label{Eq:C_theta}
C(\vartheta) \; \simeq \;
\frac 1{4\pi} \, \sum_{l=2}^\infty \, (2l+1) \, C_l \, P_l(\cos\vartheta)
\hspace{10pt} .
\end{equation}

\begin{figure}[tbh]
\begin{center}
\vspace*{-90pt}\hspace*{-90pt}\begin{minipage}{7cm}
\includegraphics[width=11.0cm]{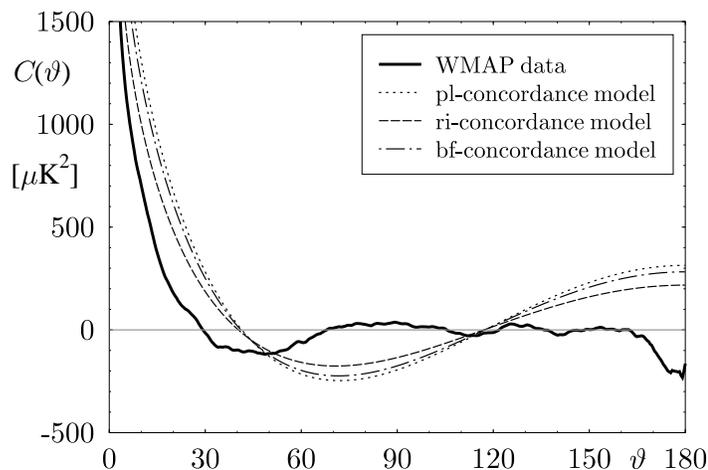}
\end{minipage}
\vspace*{-45pt}
\end{center}
\caption{\label{Fig:C_Theta_wmap}
$C(\vartheta)$ from first year WMAP data (solid curve) in comparison
to the three concordance models provided by the WMAP team
at {\sl http://lambda.gsfc.nasa.gov}:
model {\it pl} assumes a power law fit to the WMAP, CBI and ACBAR data,
model {\it ri} assumes a running index fit to the WMAP, CBI and ACBAR data,
and finally,
model {\it bf} assumes a running index fit to the WMAP, CBI and ACBAR data
and additionally takes the large scale data from 2dF and Ly-alpha into
account.
}
\end{figure}

In figure \ref{Fig:C_Theta_wmap} we show $C(\vartheta)$ as solid curve as
measured by WMAP \cite{Bennett_et_al_2003}.
The important conclusion to be drawn from figure \ref{Fig:C_Theta_wmap} is
that the temperature correlation function displays very weak correlations
at wide angles, $60^\circ \lesssim \vartheta \lesssim 160^\circ$.
The data are compared with three different versions of the concordance
model obtained by the WMAP team as best fits of their data combined
with various other data.
The model {\it pl} (dotted curve) assumes a power law fit,
i.\,e.\ with no running index, to the WMAP \cite{Bennett_et_al_2003},
CBI \cite{Mason_et_al_2003,Pearson_et_al_2003} and
ACBAR \cite{Kuo_et_al_2004} data,
the model {\it ri} (dashed curve) assumes a running index fit to
the  WMAP, CBI and ACBAR data, and finally,
the model {\it bf} (dashed-dotted curve) assumes again a running index fit
to the  WMAP, CBI and ACBAR data and additionally takes the
large scale structure data from 2dF \cite{Colless_et_al_2001} and
Lyman-$\alpha$ \cite{Croft_et_al_2002} into account.
It is seen from figure \ref{Fig:C_Theta_wmap} that the difference
between the three concordance models is not considerable.
One makes, however, the important observation that the three concordance
models using the best-fit values for the cosmological parameters
as obtained by WMAP do not reproduce the experimentally observed
suppression of power at wide angles,
leaving us with ``the mystery of the missing fluctuations''
\cite{Weeks_2004}.
In the case of the largest angles,
$160^\circ \lesssim \vartheta \le 180^\circ$,
where the WMAP curve in figure \ref{Fig:C_Theta_wmap} shows
even negative temperature correlations, one might argue
that the measurements in the nearly back-to-back configuration
are very subtle due to the need to apply a Galactic cut to
the data and, furthermore, due to geometrical reasons,
since the expectation values of $C(\vartheta)$ are computed
from a much smaller number of pixels,
and thus the discrepancy at these angles should not be overemphasized.
There is, however, also a discrepancy between the three concordance models
and the WMAP data at small angles, $\vartheta \lesssim 40^\circ$,
which is clearly visible in figure \ref{Fig:C_Theta_wmap}.

Depending on certain priors, the WMAP team reported
\cite{Spergel_et_al_2003} for the total energy density
$\Omega_{\hbox{\scriptsize tot}} = 1.02\pm0.02$ together with
$\Omega_{\hbox{\scriptsize bar}} = 0.044\pm0.004$,
$\Omega_{\hbox{\scriptsize mat}} = 0.27\pm0.04$,
and $h=0.71_{-0.03}^{+0.04}$ for the present day reduced Hubble constant
(the errors give the $1\sigma$-deviation uncertainties).
Taken at their face value, these parameters hint to a positively
curved Universe.
Recently Luminet et al.\ \cite{Luminet_Weeks_Riazuelo_Lehoucq_Uzan_2003}
studied the Poincar\'e dodecahedral space
(for details, see section \ref{Sec:spherical_space_forms})
which is one of the well-known space forms with constant positive curvature
(see also \cite{Uzan_Riazuelo_Lehoucq_Weeks_2003,Ellis_2003,Weeks_2004}).
The authors of ref.\ \cite{Luminet_Weeks_Riazuelo_Lehoucq_Uzan_2003}
computed the CMB multipoles for $l=2,3$ and 4,
fitted the overall normalization factor to match the WMAP data at $l=4$,
and then examined the prediction for $l=2$ and $l=3$ as a function of
$\Omega_{\hbox{\scriptsize tot}}$.
For $\Omega_{\hbox{\scriptsize tot}}=1.013$ they found a strong suppression
of the power at $l=2$ and weak suppression at $l=3$ in agreement
with the WMAP data.
Since the eigenfunctions of the Poincar\'e dodecahedron are not known
analytically, they must be computed numerically
\cite{Lehoucq_Weeks_Uzan_Gausmann_Luminet_2002,Lachieze_Rey_2004}.
In ref.\ \cite{Luminet_Weeks_Riazuelo_Lehoucq_Uzan_2003},
only the first three modes with wave number $\beta=13, 21$ and 25
(comprising in total 59 eigenfunctions) have been used,
which in turn restricted the discussion to the multipoles $l\le 4$.
There thus remains the question about how this extremely low wave number
cut-off affects the prediction of the multipoles,
since experience shows that increasing the cut-off usually enhances
the integrated Sachs-Wolfe contribution.

In this paper, we present a thorough discussion of the CMB anisotropy
for the dodecahedral space  topology which is based on the computation
of the first 10521 eigenfunctions corresponding to the large wave number
cut-off $\beta=155$.
Taking within the tight-coupling approximation not only the ordinary,
but also the integrated Sachs-Wolfe and also the Doppler effect into account,
we are able to predict sufficiently many multipole moments
such that a detailed comparison of the dodecahedral model
with the WMAP data can be performed.
We find that the temperature correlation function for the
dodecahedral universe possesses very weak correlations at large scales
in nice agreement with the WMAP data for $\Omega_{\hbox{\scriptsize tot}}$
in the range $1.016\dots 1.020$.
Besides the suppression of large-scale fluctuations,
the dodecahedral model predicts 6 pairs of matching circles in the sky
possessing an angular radius of $40^\circ\dots50^\circ$
for the slight curvature of space corresponding to the
cited range of $\Omega_{\hbox{\scriptsize tot}}$.

We begin in section \ref{Sec:spherical_space_forms} by 
describing the spherical space forms and, in particular, the Poincar\'e
dodecahedron, and discuss the general properties of the
vibrational modes.
In section \ref{Numerical_determination},
we outline our numerical method to determine the eigenmodes numerically.
In section \ref{angular_power_spectrum_and_correlation_function},
we calculate the angular power spectrum and the temperature
correlation function for the dodecahedral space and compare the
predictions with the WMAP data.
As a quantitative test, we study the $S$ statistic introduced
in \cite{Spergel_et_al_2003}.
For this statistic, it is found that nice agreement with the
WMAP data is obtained for 
$\Omega_{\hbox{\scriptsize tot}} = 1.016\dots 1.020$.
In section \ref{circles-in-the-sky-signature},
we study the circles-in-the-sky signature using a quantity
$\Sigma$ introduced in \cite{Cornish_Spergel_Starkman_1998b}
(called $S$ in \cite{Cornish_Spergel_Starkman_1998b}).
In the last section, there is a discussion and summary.

\section{The spherical space forms and their vibrational modes}

\label{Sec:spherical_space_forms}
The three-dimensional spaces ${\cal M}^3$ of constant positive
curvature $K=+1$ (spherical spaces) were classified by 1932
\cite{Threlfall_Seifert_1930,Threlfall_Seifert_1932}
and are given by the quotient ${\cal M}^3 = {\cal S}^3/\Gamma$
of the 3-sphere ${\cal S}^3$ under the action of a discrete fixed-point
free subgroup $\Gamma \subset \hbox{SO(4)}$ of the isometries of ${\cal S}^3$.
All these manifolds are compact and are, apart from their
universal covering ${\cal S}^3$, multi-connected.

The unit 3-sphere ${\cal S}^3$ is defined as the three-dimensional
hypersurface of the four-dimensional unit ball in flat
four-dimensional space with Cartesian coordinates
$(w,x,y,z)\in \mathbb{R}^4$, i.\,e.\ ${\cal S}^3$ is given by the set of
points which satisfy the condition $w^2 + x^2 + y^2 + z^2 \; = \; 1$.
The spatial distance $|d \vec x\,|$ between neighbouring points
is given by
\begin{equation}
\label{Eq:unit_sphere_metric}
d \vec x\,^2 \; = \; dw^2 + dx^2 + dy^2 + dz^2
\hspace{10pt} ,
\end{equation}
and the full four-dimensional space-time line element reads
\begin{equation}
\label{Eq:unit_sphere_metric_full}
ds^2 \; = \; c^2 dt^2 - R^2(t) \, d \vec x\,^2
\; = \; a(\eta)^2 \left[ d\eta^2 - d \vec x\,^2 \right]
\hspace{10pt} .
\end{equation}
Here $R(t) = a(\eta)$ denotes the cosmic scale factor as a function
of cosmic time $t$ respectively conformal time $\eta$
$(d\eta = c\,dt/R(t))$.
The metric (\ref{Eq:unit_sphere_metric_full}) has the most general form
assuming that the Universe is homogeneous and isotropic
(cosmological principle).

The condition $w^2 + x^2 + y^2 + z^2 \; = \; 1$ can be used to eliminate
one spatial coordinate, $w$ say, and one is left with the three
space variables $(x,y,z)$.
In the following, we shall also use the three angular variables
$(\chi,\theta,\phi)$ defined by
\begin{eqnarray}
\nonumber
w & = & \cos\chi
\hspace{10pt} , \hspace{10pt}
x \; = \; \sin\chi \sin\theta \cos\phi
\hspace{10pt} , \\
\label{Eq:angular_variables}
y & = & \sin\chi \sin\theta \sin\phi
\hspace{10pt} , \hspace{10pt}
z \; = \; \sin\chi \cos\theta
\hspace{10pt} ,
\end{eqnarray}
with $0\le \chi \le \pi$, $0\le \theta \le \pi$, $0\le \phi \le 2\pi$.
It is seen from (\ref{Eq:angular_variables}) that $(x,y,z)$ describe
for a fixed value of $\chi$ an ${\cal S}^2$-sphere with radius
$\sin\chi$ and volume $V^{{\cal S}^2}(\chi)=\pi(2\chi-\sin 2\chi)$,
where $\theta$ and $\phi$ play the familiar r\^ole of
polar and azimuth angle.
The ${\cal S}^2$-sphere with radius $\sin\chi$ can be considered as a
cross section of the unit ${\cal S}^3$-sphere.
Varying $\chi$ between 0 and $\pi$, one obtains an infinite sequence
of cross sections (=2-spheres) of ${\cal S}^3$,
whose radius grows from 0 to 1 (for $0\le \chi \le \frac\pi 2$),
and then shrinks again to zero (for $\frac\pi 2 \le \chi \le \pi$),
which provides a useful way to visualize the 3-sphere.

With (\ref{Eq:unit_sphere_metric}) and (\ref{Eq:angular_variables}),
the line element on ${\cal S}^3$ can be rewritten as
\begin{equation}
\label{Eq:unit_sphere_metric_angular_variables}
d \vec x\,^2 \; = \;
d\chi^2 \, + \, \sin^2\chi\,(d\theta^2 + \sin^2\theta\, d\phi^2)
\end{equation}
which has to be inserted into eq.\,(\ref{Eq:unit_sphere_metric_full}).

For a study of the large-scale anisotropies in the CMB produced by
scalar perturbations, we need the (regular) eigenmodes of the covariant
Laplacian $\Delta$ on ${\cal S}^3$,
i.\,e.\ the solutions of the Helmholtz equation
$(\Delta + E^{{\cal S}^3}) \Psi^{{\cal S}^3} = 0$
describing the vibrations on ${\cal S}^3$.
Using the coordinates $(\chi,\theta,\phi)$ and the metric
(\ref{Eq:unit_sphere_metric_angular_variables}) and separating the
variables, the normalized solutions read \cite{Schroedinger_1938,%
Schroedinger_1939,Schroedinger_1940a,Schroedinger_1940b,%
Harrison_1967,Abbott_Schaefer_1986}
$(\beta \in \mathbb{N}; l=0,1,\dots,\beta-1; m=-l,-l+1,\dots,l)$
\begin{equation}
\label{Eq:eigenmode}
\Psi_{\beta l m}^{{\cal S}^3}(\vec x\,) \; = \;
R_{\beta l}(\chi) \, \tilde Y_{lm}(\theta,\phi)
\hspace{10pt} .
\end{equation}
Here $\tilde Y_{lm}(\theta,\phi)$ are real spherical harmonics on the
unit sphere ${\cal S}^2$, and the ``radial functions'' $R_{\beta l}(\chi)$
are given by
\begin{equation}
\label{Eq:radial_function}
R_{\beta l}(\chi) \; = \;
A_{\beta l} \, (\sin\chi)^l \, C_{\beta-l-1}^{l+1}(\cos \chi)
\hspace{10pt} ,
\end{equation}
where $C_\mu^\nu(x)$ are the Gegenbauer polynomials
(which can also be expressed in terms of the associated Legendre functions
$P_\mu^\nu(x)$).
The normalization factor in (\ref{Eq:radial_function})
\begin{equation}
\label{Eq:radial_function_normalization}
A_{\beta l} \; = \;
2^{l+\frac 12} \, l! \, \sqrt{\frac\beta\pi \,
\frac{(\beta-l-1)!}{(\beta+l)!}}
\end{equation}
follows from the orthonormality relations
\begin{equation}
\label{Eq:orthonormalization}
\int_{{\cal S}^3} d\mu(\vec x\,) \,
{\Psi^{{\cal S}^3}_{\beta l m}}(\vec x\,) \,
\Psi^{{\cal S}^3}_{\beta' l' m'}(\vec x\,) \; = \;
\delta_{\beta\beta'} \, \delta_{l l'} \, \delta_{m m'}
\hspace{10pt} ,
\end{equation}
\begin{equation}
\label{Eq:orthonormalization_radial}
\int_0^\pi d\chi \, \sin^2\chi \,
R_{\beta l}(\chi) \, R_{\beta' l}(\chi) \; = \;
\delta_{\beta\beta'}
\hspace{10pt} .
\end{equation}
The volume element on ${\cal S}^3$ reads (with the standard
expression $d\Omega = \sin\theta d\theta d\phi$ on ${\cal S}^2$)
\begin{equation}
\label{Eq:volume_element}
d\mu(\vec x\,) \; = \; \sin^2\chi \, d\chi \, d\Omega
\hspace{10pt} .
\end{equation}
Notice that the radial functions $R_{\beta l}(\chi)$ are even or odd
(depending on whether $\beta-l-1$ is an even or odd integer)
about the equator $\chi=\frac\pi 2$, since they satisfy the periodic
boundary condition
\begin{equation}
\label{Eq:boundary_condition}
R_{\beta l}(\pi-\chi) \; = \; (-1)^{\beta-l-1} \, R_{\beta l}(\chi)
\hspace{10pt} .
\end{equation}
Since ${\cal S}^3$ is compact, the vibrational modes are discrete,
\begin{equation}
\label{Eq:eigenvalues}
E_\beta^{{\cal S}^3} \; = \; \beta^2 - 1
\hspace{10pt} ,
\end{equation}
where $\beta\in \mathbb{N}$ plays the r\^ole of the wave number.
The fundamental mode of ${\cal S}^3$ has an eigenvalue of
$E_1^{{\cal S}^3} = 0$ corresponding to the constant wave function
$\Psi^{{\cal S}^3}_{100} = 1/\sqrt{V^{{\cal S}^3}}$,
where $V^{{\cal S}^3} = \int_{{\cal S}^3} d\mu(\vec x\,) = 2\pi^2$
is the volume of the unit sphere ${\cal S}^3$.
Since the eigenvalues of ${\cal S}^3$ do not depend on $l$ and $m$,
the modes are degenerate, and the multiplicity of the mode $\beta$
is given by
\begin{equation}
\label{Eq:multiplicity}
r^{{\cal S}^3}(\beta) \; = \; \sum_{l=0}^{\beta-1} (2 l + 1) \; = \;
\beta^2
\hspace{10pt} .
\end{equation}
If $N^{{\cal S}^3}(k)$ denotes the number of the vibrational modes on
${\cal S}^3$ with wave number smaller than or equal to $k\in\mathbb{N}$,
we obtain
$$
N^{{\cal S}^3}(k) \; = \;
\sum_{\beta=1}^k r^{{\cal S}^3}(\beta) \; = \;
\frac 13\, k^3 \, + \, \frac 12\, k^2 \, + \, \frac 16\, k \; = \;
\frac{V^{{\cal S}^3}}{6\pi^2} \, k^3 \, + \, O(k^2)
$$
in agreement with Weyl's asymptotic law for $k\to\infty$.

We now turn to the multi-connected spherical spaces
${\cal S}^3/\Gamma$, $\Gamma\subset \hbox{SO(4)}$,
possessing the volume $V({\cal S}^3/\Gamma) = V^{{\cal S}^3}/N$,
where $N$ denotes the order of the group $\Gamma$.
(For reviews, see
\cite{Threlfall_Seifert_1930,Threlfall_Seifert_1932,Wolf_1974,%
Thurston_1997,Gausmann_Lehoucq_Luminet_Uzan_Weeks_2001}).
To define the discrete fixed-point free subgroups
$\Gamma \subset \hbox{SO(4)}$ of the isometries of ${\cal S}^3$,
it is advantageous to work with Hamilton quaternions
$q:= w+x\i + y\j + z\i\j$, $(w,x,y,z)\in \mathbb{R}^4$,
which are spanned by the 4 basis quaternions $\{1,\i,\j,\i\j\}$,
with the multiplication defined by $\i^2=\j^2=-1$, $\i\j=-\j\i$
plus the property that $\i$ and $\j$ commute with every real number.
The multiplication of quaternions is associative,
but not commutative.
If $q^\star := w-x\i - y\j - z\i\j$ denotes the quaternion conjugate to $q$,
the norm of $q$ is defined by $|q|^2 := q q^\star = w^2 + x^2 + y^2 + z^2$.
Quaternions with norm $1$ are called unit quaternions.
For a unit quaternion $q$, there exists an inverse $q^{-1}$
which is simply given by $q^{-1}=q^\star$.
It is obvious that the unit  3-sphere ${\cal S}^3$ can be identified with
the multiplicative group of unit quaternions.
The distance $d(q_1, q_2)$ between any two points $q_1$ and $q_2$ on
${\cal S}^3$ is given by
$\cos d(q_1, q_2) = w_1w_2 + x_1x_2 + y_1y_2 + z_1z_2$
and is a point-pair invariant,
i.\,e.\ $d(\gamma q_1,\gamma q_2) = d(q_1,q_2)$
$\forall \gamma \in \hbox{SO(4)}$.

The group SO(4) is isomorphic to
${\cal S}^3\times{\cal S}^3/\{\pm(1,1)\}$,
the two factors corresponding to the left and right group actions.
In the following, we are only interested in subgroups
$\Gamma\in\hbox{SO(4)}$ which lead to homogeneous manifolds
${\cal S}^3/\Gamma$.
Such groups possess only so-called right-handed Clifford translations.
A quaternion $\gamma\in\Gamma$ corresponds to a Clifford translation,
if it translates all points $q_1,q_2\in {\cal S}^3$
by the same distance $\chi$,
i.\,e.\ $d(q_1,\gamma q_1) = d(q_2,\gamma q_2) = \chi$.
The right-handed Clifford transformations $\gamma_k\in\Gamma$ in
${\cal S}^3$ are defined by left-multiplication of an arbitrary unit
quaternion $q \in {\cal S}^3$ by
$\gamma_k := a_k + b_k\i + c_k\j + d_k \i\j$, i.\,e.\
\begin{eqnarray}
q \; \mapsto \gamma_k \, q & = & \nonumber
\left( a_k + b_k\i + c_k\j + d_k \i\j \right) \, \cdot \,
\left( w+x\i + y\j + z\i\j \right) \\
& = & \nonumber
\left(a_k w - b_k x - c_k y - d_k z \right)
\\ & \,+ & \nonumber
\left(b_k w + a_k x - d_k y + c_k z \right) \, \i
\\ & \,+ & \label{Eq:right-handed-Clifford}
\left(c_k w + d_k x + a_k y - b_k z \right) \, \j
\\ & \,+ & \nonumber
\left(d_k w - c_k x + b_k y + a_k z \right) \, \i \j
\hspace{10pt} .
\end{eqnarray}
The right-handed Clifford translation (\ref{Eq:right-handed-Clifford})
acts as a right-handed corkscrew fixed-point free rotation of ${\cal S}^3$.

For our numerical computations of the eigenfunctions,
it is convenient to represent the action (\ref{Eq:right-handed-Clifford})
of a group element $\gamma_k\in\Gamma$ by the following orthogonal
$4\times 4$ matrix $M_k$
\begin{equation}
\label{Eq:matrix_right-handed-Clifford_}
M_k \; = \; \left( \begin{tabular}{rrrr}
$a_k$ &$ -b_k$ &$ -c_k$ &$ -d_k$ \\
$b_k$ &$  a_k$ &$ -d_k$ &$  c_k$ \\
$c_k$ &$  d_k$ &$  a_k$ &$ -b_k$ \\
$d_k$ &$ -c_k$ &$  b_k$ &$  a_k$ \end{tabular} \right)
\hspace{10pt} ,
\end{equation}
which acts from the left in an obvious way on the basis quaternions
$(1,\i,\j,\i\j)$.

The following groups lead to homogeneous manifolds ${\cal S}^3/\Gamma$
\cite{Threlfall_Seifert_1930,Threlfall_Seifert_1932,Wolf_1974,Thurston_1997}:
\begin{itemize}
\item The cyclic groups $Z_n$ of order $n$ $(n\ge 1)$.
\item The binary dihedral groups $D_m^\star$ of order $4m$ $(m\ge 2)$.
\item The binary tetrahedral group $T^\star$ of order $24$.
\item The binary octahedral group $O^\star$ of order $48$.
\item The binary icosahedral group $I^\star$ of order $120$.
\end{itemize}

Let us consider the binary icosahedral group $I^\star$ in more detail
consisting of $N=120$ group elements.
They are generated by the two right-handed Clifford translations
\cite{Gausmann_Lehoucq_Luminet_Uzan_Weeks_2001}
\begin{eqnarray}\nonumber
\gamma_1 &:& q \; \mapsto \; \gamma_1\, q \; := \; \j q \; = \;
-y + z\i + w\j - x\i\j \\ \nonumber
\gamma_2 &:& q \; \mapsto \; \gamma_2\, q
\end{eqnarray}
with the unit quaternion
$\gamma_2 := \frac \sigma 2 + \frac 1{2\sigma} \i + \frac 12\j$
corresponding to the angles
$(\chi_2,\theta_2,\phi_2) = \big( \frac\pi 5,\frac\pi 2,\arctan\sigma\big)$,
where $\sigma = \frac{\sqrt 5+1}2$ is the golden ratio.
The fundamental cell (Dirichlet domain) of the binary polyhedral space
${\cal D} := {\cal S}^3/I^\star$ is a regular dodecahedron and is
known as the {\it Poincar\'e dodecahedral space},
which is made of 12 pentagons, 30 edges and 20 corners.
120 dodecahedra tessellate the unit 3-sphere ${\cal S}^3$.
The shortest translation distance in ${\cal D}$ is given by
$d(q,\gamma_2 q)=\chi_2=\frac\pi 5$,
while $\gamma_1$ translates all points already to the much larger distance
$\chi_1=\frac\pi 2$.

The vibrations on the general homogeneous spherical 3-spaces
${\cal M}^3 = {\cal S}^3/\Gamma$ are determined again by the
Helmholtz equation on ${\cal S}^3$, but now with the periodic
boundary conditions corresponding to the defining fundamental cells
(polyhedra).
Thus the eigenfunctions $\Psi^{{\cal M}^3}(\vec x\,)$ on ${\cal M}^3$
can be expanded into the eigenfunctions $\Psi^{{\cal S}^3}(\vec x\,)$:
\begin{equation}
\label{Eq:eigenfunction_M3}
\Psi^{{\cal M}^3,i}_\beta(\vec x\,) \; = \;
\sum_{l=0}^{\beta-1} \sum_{m=-l}^l \xi_{\beta l m}^i({\cal M}^3) \,
\Psi^{{\cal S}^3}_{\beta l m}(\vec x\,)
\hspace{10pt} ,
\end{equation}
with $E^{{\cal M}^3} = \beta^2 - 1$, 
$\beta\in \mathbb{N}$, $i=1,\dots,r^{{\cal M}^3}(\beta)$.
It is important to notice that for a given manifold ${\cal M}^3$
the wave numbers $\beta$ do not take all values in $\mathbb{N}$.
The allowed $\beta$ values together with their multiplicities
$r^{{\cal M}^3}(\beta)$ are, however, explicitly known \cite{Ikeda_1995}.
The full information of the non-trivial topology of ${\cal M}^3$
is contained in the real expansion coefficients
$\{\xi_{\beta l m}^i({\cal M}^3)\}$
which by virtue of (\ref{Eq:orthonormalization}) are given by
\begin{equation}
\label{Eq:eigenfunction_coefficient}
\xi_{\beta l m}^i({\cal M}^3) \; = \;
\int_{{\cal S}^3} d\mu(\vec x\,) \,
\Psi^{{\cal S}^3}_{\beta l m}(\vec x\,) \,
\Psi^{{\cal M}^3,i}_\beta(\vec x\,)
\hspace{10pt} .
\end{equation}
They satisfy the normalization condition
\begin{equation}
\label{Eq:eigenfunction_normalization}
\sum_{l=0}^{\beta-1} \sum_{m=-l}^l
\left( \xi_{\beta l m}^i({\cal M}^3) \right)^2 \; = \; N
\end{equation}
due to the orthonormality relation
\begin{eqnarray}
\nonumber
& & \hspace{-60pt} \int_{{\cal M}^3} d\mu(\vec x\,) \,
\Psi^{{\cal M}^3,i}_{\beta}(\vec x\,) \,
\Psi^{{\cal M}^3,i'}_{\beta'}(\vec x\,) \\
\label{Eq:eigenfunction_orthonormalization}
\hspace{40pt} & = &
\frac 1N \int_{{\cal S}^3} d\mu(\vec x\,) \,
\Psi^{{\cal M}^3,i}_{\beta}(\vec x\,) \,
\Psi^{{\cal M}^3,i'}_{\beta'}(\vec x\,) \; = \;
\delta_{\beta\beta'} \, \delta_{i i'}
\hspace{10pt} .
\end{eqnarray}
Summation of the relation (\ref{Eq:eigenfunction_normalization})
over the degeneracy index $i$ yields
\begin{equation}
\label{Eq:eigenfunction_sum_degeneracy}
\sum_{l=0}^{\beta-1} \sum_{m=-l}^l  \sum_{i=1}^{r^{{\cal M}^3}(\beta)}
\left( \xi_{\beta l m}^i({\cal M}^3) \right)^2 \; = \;
N \, r^{{\cal M}^3}(\beta)
\hspace{10pt} .
\end{equation}
In section \ref{Numerical_determination} we shall determine the
coefficients $\xi_{\beta l m}^i({\cal M}^3)$ numerically up to some
wave number cut-off $\beta_{\hbox{\scriptsize max}}$.
Here we state only the following \underline{Conjecture}
for homogeneous space forms $(0\le l \le \beta-1)$
\begin{equation}
\label{Eq:Conjecture}
\frac 1{2l+1} \sum_{m=-l}^l  \sum_{i=1}^{r^{{\cal M}^3}(\beta)}
\left( \xi_{\beta l m}^i({\cal M}^3) \right)^2 \; = \;
N \, \frac{r^{{\cal M}^3}(\beta)}{\beta^2}
\hspace{10pt} ,
\end{equation}
for which a numerical check will be discussed below.
Let us also note that in our application to the CMB anisotropy,
we only need the modes with $\beta \ge 3$ (if they exist),
since the values $\beta=1,2$ correspond to modes
which are pure gauge terms \cite{Bardeen_1980}.

\section{Numerical determination of the eigenmodes}

\label{Numerical_determination}

For the homogeneous 3-manifolds ${\cal M}^3={\cal S}^3/\Gamma$,
which we are considering here,
the eigenvalues of the Helmholtz equation are given by
$E^{{\cal M}^3}_\beta = \beta^2 -1$, $\beta\in\mathbb{N}$,
and even their multiplicities $r^{{\cal M}^3}(\beta)$
are explicitly known \cite{Ikeda_1995}.
As already mentioned in section \ref{Sec:spherical_space_forms},
the wave number $\beta$ does not assume all values in $\mathbb{N}$,
and thus the multiplicity $r^{{\cal M}^3}(\beta)$ refers only to
the $\beta$ values which appear in the spectrum.
E.\,g.\ in the case of the Poincar\'e dodecahedral space ${\cal D}$,
there exist no eigenmodes for even $\beta$-values, and, furthermore,
also among the odd values $\beta\in2\mathbb{N}+1$ there are finitely many gaps,
i.\,e.\ $\beta$ values for which no eigenmode exists.
Explicitly, one has \cite{Ikeda_1995}
\begin{equation}
\label{Eq:E_beta}
E_\beta^{\cal D} \; = \; \beta^2 \, - \, 1
\end{equation}
with $\beta \in \{1,13,21,25,31,33,37,41,43,45,49,51,53,55,57\} \cup
\{2n+1, n \ge 30\}$ and
\begin{equation}
\label{Eq:Ikeda}
r^{\cal D}(\beta) \; = \;
\beta\left( \left[\frac{\beta-1}{10}\right] \, + \,
\left[\frac{\beta-1}6\right] \, + \,
\left[\frac{\beta-1}4\right] \, - \,
\frac{\beta-3}2 \right)
\hspace{10pt} .
\end{equation}

The fact that the eigenvalues respectively the wave numbers $\beta$
for a given manifold ${\cal M}^3$ are explicitly known,
facilitates greatly the numerical computation of the eigenfunctions
$\Psi_\beta^{{\cal M}^3,i}(\vec x\,)$,
since it saves us from the time-consuming numerical search for the
allowed $\beta$-values as it has to be carried out,
e.\,g.\ for hyperbolic manifolds in the case of a negatively curved universe
\cite{Aurich_Steiner_1991,Aurich_Marklof_1996}.
Furthermore, since the expansion (\ref{Eq:eigenfunction_M3}) in terms
of the eigenfunctions ${\cal S}^3$ involves for a given wave number
only a finite basis of dimension $\beta^2$
(see eq.\,(\ref{Eq:multiplicity})),
the number of the real coefficients $\{\xi_{\beta l m}^i({\cal M}^3)\}$
is also restricted to $\beta^2$ for a given eigenfunction with
wave number $\beta$ and degeneracy index $i$.

The eigenfunctions have to satisfy the fundamental periodicity condition
\begin{equation}
\label{Eq:periodicity_condition}
\Psi_\beta^{{\cal M}^3,i}(\gamma_k q) \; = \;
\Psi_\beta^{{\cal M}^3,i}(q)
\hspace{10pt} , \hspace{10pt}
\forall \; q \in {\cal S}^3
\hspace{10pt} , \hspace{10pt}
\forall \; \gamma_k \in \Gamma
\hspace{10pt} ,
\end{equation}
where $\gamma_k$ denotes a group element of the group
$\Gamma \subset \hbox{SO(4)}$
which defines the manifold ${\cal M}^3={\cal S}^3/\Gamma$.
We will use the condition (\ref{Eq:periodicity_condition}) in a
collocation algorithm.
Thus, the condition (\ref{Eq:periodicity_condition}) will be
imposed on $L>\beta^2$ randomly distributed points $q$ on ${\cal S}^3$.
To generate such random points,
we use instead of the coordinates $(w,x,y,z)$ the coordinates
$(v,\alpha,\gamma)$ defined by
\begin{equation}
\label{Eq:trafo_random_points}
\begin{tabular}{rcl}
$w$ & $=$ & $\sqrt{1-2v} \, \cos\alpha$ \\
$x$ & $=$ & $\sqrt{2v} \, \cos\gamma$ \\
$y$ & $=$ & $\sqrt{2v} \, \sin\gamma$ \\
$z$ & $=$ & $\sqrt{1-2v} \, \sin\alpha$
\end{tabular}
\end{equation}
with $v\in[0,\frac 12]$ and $\alpha, \gamma \in [0,2\pi]$.
The line element (\ref{Eq:unit_sphere_metric}) takes then the form
$$
d\vec x\,^2 \; = \;
\frac{dv^2}{2v(1-2v)} \, + \, (1-2v) \, d\alpha^2 \, + \, 2 v d\gamma^2
\hspace{10pt} ,
$$
which gives the volume element $d\mu = dv \, d\alpha \, d\gamma$.
The random points $q$ on ${\cal S}^3$ are now generated
by randomly choosing the coordinates $(v,\alpha,\gamma)$ and
then calculating from (\ref{Eq:trafo_random_points}) the
coordinates $(w,x,y,z)$ of the corresponding unit quaternion $q$.
In this way, we generate $L$ random points $\{q_1,q_2,\dots,q_L\}$.
An arbitrary group element $\gamma_k\in\Gamma$ maps the point $q_r$
to the point $\tilde q_r := \gamma_k q_r$ on ${\cal S}^3$,
where in the following computation the group elements $\gamma_k$
run over $\Gamma\backslash\{1\}$.
The action of $\gamma_k$ will be represented by the matrix $M_k$,
see eqs.\,(\ref{Eq:right-handed-Clifford},
\ref{Eq:matrix_right-handed-Clifford_}).
After having generated in this manner $L$ mirror points
$\{\tilde q_1,\tilde q_2,\dots,\tilde q_L\}$,
one imposes the condition (\ref{Eq:periodicity_condition}) at all
those $L$ points.
Inserting the expansion (\ref{Eq:eigenfunction_M3}),
one finally obtains the following over-determined system of equations
$(r=1,\dots,L>\beta^2)$ for the $\beta^2$ coefficients
$\{\xi_{\beta l m}^i({\cal M}^3)\}\,$:
\begin{equation}
\label{Eq:coeff_system}
\sum_{l=0}^{\beta-1} \sum_{m=-l}^l \, \left[
\Psi_{\beta l m}^{{\cal S}^3}(q_r) \, - \,
\Psi_{\beta l m}^{{\cal S}^3}(\tilde q_r) \right] \,
\xi_{\beta l m}^i({\cal M}^3) \; = \; 0
\hspace{10pt} .
\end{equation}
The system (\ref{Eq:coeff_system}) is for a given $\beta$-value
of the form
$\sum_{s=1}^{\beta^2} B_{rs} \xi_s = 0$, $r=1,\dots,L$,
where the $(L \times \beta^2)$-matrix $B=B(\beta)$
is given by the square bracket in eq.\,(\ref{Eq:coeff_system})
and is thus completely determined in terms of the eigenfunctions
on the covering space ${\cal S}^3$.
The vector $\xi=\xi(\beta)$ has the $\beta^2$ components
$\xi_s \equiv \xi_{\beta l m}({\cal M}^3)$,
i.\,e.\ $s$ denotes the multi-index
$\{00, 1-1, 10, 11,\dots,\beta-1\, \beta-1\}$.
The system (\ref{Eq:coeff_system}) is numerically solved by using
the singular value decomposition.
One then obtains $r^{{\cal M}^3}(\beta)$ linearly independent solutions
for the expansion coefficients $\{\xi_{\beta l m}^i({\cal M}^3)\}$
which are distinguished by the index $i$.
In a final step, the coefficients are normalized according to
the condition (\ref{Eq:eigenfunction_normalization}).

Using this method, we have computed the expansion coefficients for
$\Gamma = I^\star$, i.\,e.\ for the Poincar\'e dodecahedral space
${\cal D}= {\cal S}^3/I^\star$, for $\beta \le 155$
comprising the first 10\,521 eigenfunctions.
The normalized coefficients have been used to check the
conjecture (\ref{Eq:Conjecture}) which has been found to hold
within a numerical accuracy of 13 digits.
In addition, we have computed the coefficients
$\{\xi_{\beta l m}^i({\cal M}^3)\}$ and
checked the conjecture (\ref{Eq:Conjecture})
for $\beta\leq 33$ for some cyclic groups, some dihedral groups,
the binary tetrahedral and binary octahedral group and have found again
excellent numerical agreement.
However, for the inhomogeneous lens spaces
\cite{Gausmann_Lehoucq_Luminet_Uzan_Weeks_2001}
$L(12,5)$ and $L(72,17)$, we have found
that the conjecture (\ref{Eq:Conjecture}) does not hold.
We thus conclude that the conjecture (\ref{Eq:Conjecture})
is only valid for the homogeneous 3-spaces.

\section{The angular power spectrum $\delta T_l^2$ and the
correlation function $C(\vartheta)$}

\label{angular_power_spectrum_and_correlation_function}
In order to determine the angular power spectrum $\delta T_l^2$
and the temperature correlation function $C(\vartheta)$,
one has to compute the temperature fluctuation
$\frac{\delta T}T(\hat n)$.
This is done using the Sachs-Wolfe formula which is given
within the tight-coupling approximation by $(c=1)$
\begin{eqnarray} \nonumber \hspace{-25pt}
\frac{\delta T}T(\hat n) & = &
\int d^3k \; \left[ \left( \Phi_{\vec k}(\eta) +
\frac{\delta_{\vec k,\gamma}(\eta)}4 +
\frac{a(\eta) V_{\vec k,\gamma}(\eta)}{E_{\vec k}}
\frac{\partial}{\partial \tau} \right)
\Psi_{\vec k}^{{\cal M}^3}(\tau(\eta),\theta,\phi)
\right]_{\eta=\eta_{\hbox{\scriptsize{SLS}}}}
\\ & & \hspace{10pt}
\label{Eq:Sachs_Wolfe_tight_coupling}
\, + \,
2 \int d^3k \; \int_{\eta_{\hbox{\scriptsize{SLS}}}}^{\eta_0} d\eta \,
\frac{\partial\Phi_{\vec k}(\eta)}{\partial\eta} \,
\Psi_{\vec k}(\tau(\eta),\theta,\phi)
\hspace{10pt} ,
\end{eqnarray}
where the $\vec k$-integration has to be replaced by a summation
over $\beta$ in the case of a discrete spectrum.
(Here $\eta_{\hbox{\scriptsize SLS}}$ denotes the conformal time
at recombination corresponding to a redshift
$z_{\hbox{\scriptsize SLS}}=1089$.)
$\Psi_{\vec k}^{{\cal M}^3}$ are the eigenfunctions on ${\cal M}^3$,
and $\Phi_{\vec k}(\eta)$ are the expansion coefficients of the
metric perturbation $\Phi$, see eq.\,(\ref{Eq:metric_perturbed}) below.
$(\tau(\eta),\theta,\phi)$ denote the spherical coordinates of the photon
path in the direction $\hat n$,
where we now assume that the observer is at the origin of the
coordinate system,
i.\,e.\ at $(\chi_{\hbox{\scriptsize obs}},\theta_{\hbox{\scriptsize obs}},
\phi_{\hbox{\scriptsize obs}}) = (0,0,0)$.
Here, $\delta_{\vec k,\gamma}(\eta)$ is the expansion
coefficient of the relative perturbation in the radiation component, and
$V_{\vec k,\gamma}(\eta)$ is the expansion
coefficient of the spatial covariant divergence
of the velocity field of the tightly coupled radiation-baryon components.
The method by which the quantities occurring in
(\ref{Eq:Sachs_Wolfe_tight_coupling}) are numerically computed,
is described in detail in Section 2 of
\cite{Aurich_Lustig_Steiner_Then_2004a}.

The ordinary Sachs-Wolfe (SW) contribution to the
temperature fluctuation $\delta T(\hat n)$
is a combination of the gravitational redshift due to the
gravitational potential $\Phi(\eta,\tau,\theta,\phi)$ at the
surface of last scattering (SLS) and the intrinsic temperature
fluctuation $\frac 14 \delta_\gamma$ due to the imposed
entropic initial conditions.
Since the SW contribution,
i.\,e.\ the first two terms in the brackets in
(\ref{Eq:Sachs_Wolfe_tight_coupling}) (see also eq.\,(\ref{Eq:NSW}) below),
gives a dominant contribution to the CMB anisotropies at large scales,
we first present an exact analytic expression for the expansion coefficients
$\{a_{lm}\}$,
the multipole moments $\{C_l\}$ and the temperature correlation
function $C(\vartheta)$ for the Poincar\'e dodecahedral universe
taking only the scalar perturbations into account due to the SW effect.

For an energy-momentum tensor $T_{\mu\nu}$
with $T_{ij}=0$ for $i\neq j$ and $i,j=1,2,3$, the line element
can in conformal Newtonian gauge be written as \cite{Bardeen_1980}
\begin{equation}
\label{Eq:metric_perturbed}
ds^2 \; = \;
a^2(\eta) \, \left[ (1+2\Phi) d\eta^2 - (1-2\Phi) |d\vec x\,|^2 \,\right]
\hspace{10pt} .
\end{equation}
Denoting by $\hat n = \hat n(\theta,\phi)$ the unit vector in the direction
in which the photons are observed, the SW formula reads
\begin{eqnarray}
\label{Eq:NSW}
\frac{\delta T^{\hbox{\scriptsize SW}}(\hat n)}T & =&
\Phi(\eta_{\hbox{\scriptsize SLS}},\tau_{\hbox{\scriptsize SLS}},\theta,\phi)
\, + \,
\frac 14 \delta_\gamma(\eta_{\hbox{\scriptsize SLS}},
\tau_{\hbox{\scriptsize SLS}},\theta,\phi)
\\ & \simeq &
\label{Eq:NSW_approx}
\frac 13 \Phi(\eta_{\hbox{\scriptsize SLS}},
\tau_{\hbox{\scriptsize SLS}},\theta,\phi)
\end{eqnarray}
with $\tau_{\hbox{\scriptsize SLS}} := \eta_0 - \eta_{\hbox{\scriptsize SLS}}$.
(Here $\eta_0$ denotes the conformal time at the present epoch.)
The SW contribution (\ref{Eq:NSW}) can be approximated as in
(\ref{Eq:NSW_approx}) for modes well outside the horizon at
the time $\eta_{\hbox{\scriptsize SLS}}$ of last scattering,
i.\,e.\ for modes with $\beta \ll \frac{2\pi}{\eta_{\hbox{\scriptsize SLS}}}$.
The approximation (\ref{Eq:NSW_approx}) is used in the following 
derivation of the analytic expressions for
$C_l^{\hbox{\scriptsize SW}}$, eq.\,(\ref{Eq:multipol_NSW}), and
$C^{\hbox{\scriptsize SW}}(\vartheta)$, eq.\,(\ref{Eq:C_theta_NSW}).
In the actual numerical computation,
we use (\ref{Eq:Sachs_Wolfe_tight_coupling}) as discussed above.
For the dodecahedron ${\cal D}$, the metric perturbation can be written
as an expansion in the eigenfunctions of ${\cal D}$
\begin{equation}
\label{Eq:grav_potential}
\Phi(\eta,\tau,\theta,\phi) \; = \;
{\sum_{\beta\ge 13}} '\; \sum_{i=1}^{r^{\cal D}(\beta)}
\Phi_\beta^i(\eta) \, \Psi_\beta^{{\cal D},i}(\tau,\theta,\phi)
\hspace{10pt} .
\end{equation}
Here the prime in the summation over the modes $\beta$ indicates
that $\beta \in 2\mathbb{N}+1$ does not assume all values.
The functions $\Phi_\beta^i(\eta)$ determine the time-evolution and
will be factorized
\begin{equation}
\label{Eq:time_evolution}
\Phi_\beta^i(\eta) \; = \; \Phi_\beta^i(0) \, g_\beta(\eta)
\end{equation}
with $g_\beta(0)=1$.
The functions $g_\beta(\eta)$ do not depend on $i$,
since the associated differential equation depends only on
the eigenvalue $E_\beta^{\cal D}$ which is independent of $i$.
The initial values $\Phi_\beta^i(0)$ play the r\^ole of the primordial
fluctuation amplitudes and are Gaussian random variables with
zero expectation value and covariance
\begin{equation}
\label{Eq:covariance}
\left< \Phi_\beta^i(0)\, \Phi_{\beta'}^{i'}(0) \right> \; = \;
\delta_{\beta\beta'} \, \delta_{i i'} \, P_\Phi(\beta)
\hspace{10pt} .
\end{equation}
Here $P_\Phi(\beta)$ denotes the primordial power spectrum
which determines the weight by which the primordial modes $\beta$
are excited, on average.
The average $\left<\dots\right>$ in (\ref{Eq:covariance}) denotes an
ensemble average over the primordial perturbations.
In the following, we shall assume the scale-invariant
Harrison-Zel'dovich spectrum
\begin{equation}
\label{Eq:Harrison_Zeldovich}
P_\Phi(\beta) \; = \; \frac \alpha{\beta(\beta^2-1)}
\hspace{10pt} .
\end{equation}
The only free parameter in (\ref{Eq:Harrison_Zeldovich}) is the normalization
constant $\alpha$ which will be determined from the CMB data.

Inserting the expansion (\ref{Eq:grav_potential}) into the
Sachs-Wolfe formula (\ref{Eq:NSW_approx}) and using the eigenfunctions of the
dodecahedron (see eqs.\,(\ref{Eq:eigenfunction_M3}), (\ref{Eq:eigenmode})
and (\ref{Eq:delta_T_expansion})),
we obtain for the expansion coefficients of the SW contribution $(l\ge 2)$
\begin{equation}
\label{Eq:expansion_NSW}
a_{lm}^{\hbox{\scriptsize SW}}({\cal D}) \; = \;
\frac 13 {\sum_{\beta> l}}'\; \sum_{i=1}^{r^{\cal D}(\beta)}
\Phi_\beta^i(\eta_{\hbox{\scriptsize SLS}}) \,
\xi_{\beta l m}^i({\cal D}) \, R_{\beta l}(\tau_{\hbox{\scriptsize SLS}})
\hspace{10pt} ,
\end{equation}
which in turn yields with the definition (\ref{Eq:C_l})
and the Gaussian hypothesis (\ref{Eq:covariance}) the mean values of
the multipole moments $(l\ge 2)$
\begin{equation}
\label{Eq:multipol_NSW_} 
\hspace*{-60pt}
C_l^{\hbox{\scriptsize SW}}({\cal D}) \; = \;
\frac 19 {\sum_{\beta>l}}' \;
P_\Phi(\beta) \, g_\beta^2(\eta_{\hbox{\scriptsize SLS}}) \,
\left[ \frac 1{2l+1} \sum_{m=-l}^l \sum_{i=1}^{r^{\cal D}(\beta)}
\left(\xi_{\beta l m}^i({\cal D})\right)^2 \right] \,
R_{\beta l}^2(\tau_{\hbox{\scriptsize SLS}})
\hspace{5pt} .
\end{equation}
At this point we can use our Conjecture (\ref{Eq:Conjecture})
with $N=120$ and arrive at the final form
\begin{equation}
\label{Eq:multipol_NSW}
C_l^{\hbox{\scriptsize SW}}({\cal D}) \; = \;
\frac{40}3 {\sum_{\beta>l}}' \;
\frac{r^{\cal D}(\beta) P_\Phi(\beta)}{\beta^2} \,
g_\beta^2(\eta_{\hbox{\scriptsize SLS}}) \,
R_{\beta l}^2(\tau_{\hbox{\scriptsize SLS}})
\hspace{10pt} .
\end{equation}
The last expression shows that the lowest multipoles are suppressed
due to the discrete spectrum of the vibrational modes of ${\cal D}$.
Since the lowest modes are at $\beta=13$, 21, 25 and 31,
the mode sum starts at $\beta=13$ for the multipoles $l=2,\dots,12$,
while for the multipoles $l=13,\dots,20$ it starts only at $\beta=21$,
for the multipoles $l=21,\dots,24$ only at $\beta=25$, and
for the multipoles $l=25,\dots,30$ only at $\beta=31$.

\begin{figure}[tbh]
\begin{center}
\vspace*{-90pt}\hspace*{-90pt}\begin{minipage}{7cm}
\includegraphics[width=11.0cm]{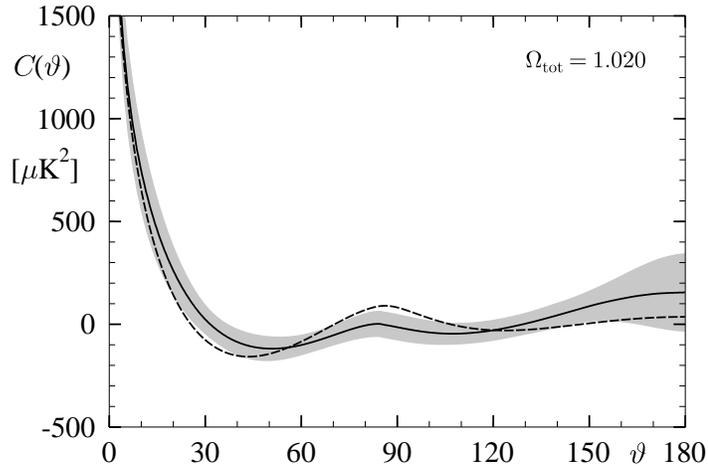}
\end{minipage}
\vspace*{-45pt}
\end{center}
\caption{\label{Fig:C_Theta_NSW}
The SW contribution (\ref{Eq:C_theta_NSW}) of the temperature correlation
function $C(\vartheta)$ is shown as a dashed curve for the
dodecahedral space ${\cal D}$ for $\Omega_{\hbox{\scriptsize tot}}=1.020$.
The solid curve shows the full result including the integrated
Sachs-Wolfe and Doppler contribution.
The $1\sigma$ deviation is shown as a grey band.
}
\end{figure}

From (\ref{Eq:multipol_NSW}) we can compute the SW contribution
to the correlation function (\ref{Eq:C_theta}) for
the Poincar\'e dodecahedron ${\cal D}$
\begin{equation}
C^{\hbox{\scriptsize SW}}(\vartheta) \; = \;
\frac{10}{3\pi} {\sum_{\beta\ge 13}}' \;
\frac{r^{\cal D}(\beta) P_\Phi(\beta)}{\beta^2} \,
g_\beta^2(\eta_{\hbox{\scriptsize SLS}}) \, \overline{C}(\beta,\vartheta)
\hspace{10pt} ,
\end{equation}
where the angular dependence is completely given by
$(P_l(\cos\vartheta) = C_l^{1/2}(\cos\vartheta))$
\begin{eqnarray}
\label{Eq:C_bar_NSW}
\overline{C}(\beta,\vartheta) & := & \nonumber
\sum_{l=2}^{\beta-1} (2l+1) \,
R_{\beta l}^2(\tau_{\hbox{\scriptsize SLS}}) \, P_l(\cos\vartheta) \\
& = &
\sum_{l=2}^{\beta-1} (2l+1) \, A_{\beta l}^2
(\sin\tau_{\hbox{\scriptsize SLS}})^{2l} \,
\left[ C_{\beta-l-1}^{l+1}(\cos\tau_{\hbox{\scriptsize SLS}})\right]^2 \,
C_l^{1/2}(\cos\vartheta)
\end{eqnarray}
after having inserted the radial function (\ref{Eq:radial_function}).
With the help of the addition theorem for the Gegenbauer polynomials
(see p.\,223 in \cite{Magnus_Oberhettinger_Soni_1966}),
one obtains for the mean value of the SW contribution to the
correlation function for the Poincar\'e dodecahedron universe
\begin{eqnarray}
C^{\hbox{\scriptsize SW}}(\vartheta) & = & \nonumber
\frac{20}{3\pi^2} {\sum_{\beta\ge 13}}' \;
\frac{r^{\cal D}(\beta) P_\Phi(\beta)}{\beta} \,
g_\beta^2(\eta_{\hbox{\scriptsize SLS}}) \,
\frac{\sin(\beta \gamma(\vartheta))}{\sin\gamma(\vartheta)} \\
& & \hspace*{85pt} \label{Eq:C_theta_NSW}
- \, \frac 1{4\pi} \, \hat C_0^{\hbox{\scriptsize SW}} \, - \,
\frac 3{4\pi} \, C_1^{\hbox{\scriptsize SW}} \, \cos\vartheta
\hspace{10pt} ,
\end{eqnarray}
where the relation
$C_{\beta-1}^1(\cos\gamma) = \frac{\sin(\beta\gamma)}{\sin\gamma}$
has been used.
The angle $\gamma=\gamma(\vartheta) \in [0,2\tau_{\hbox{\scriptsize SLS}}]$
is given by
\begin{equation}
\gamma(\vartheta) \; = \; \arccos\big( \cos^2\tau_{\hbox{\scriptsize SLS}}
+ \sin^2\tau_{\hbox{\scriptsize SLS}} \cos\vartheta \big)
\hspace{10pt} ,
\end{equation}
and the monopole and the dipole moment read
\begin{eqnarray}
\hat C_0^{\hbox{\scriptsize SW}} & = &
\frac{80}{3\pi} {\sum_{\beta\ge 13}}' \;
\frac{r^{\cal D}(\beta) P_\Phi(\beta)}{\beta^2} \,
g_\beta^2(\eta_{\hbox{\scriptsize SLS}}) \,
\left[\frac{\sin(\beta \tau_{\hbox{\scriptsize SLS}})}
{\sin\tau_{\hbox{\scriptsize SLS}}}\right]^2
\\ 
C_1^{\hbox{\scriptsize SW}} & = &
\frac{320}{3\pi} \, \sin^2(\tau_{\hbox{\scriptsize SLS}})
{\sum_{\beta\ge 13}}' \,
\frac{r^{\cal D}(\beta) P_\Phi(\beta)}{\beta^2(\beta^2-1)} \,
g_\beta^2(\eta_{\hbox{\scriptsize SLS}}) \,
\left[C_{\beta-2}^2(\cos\tau_{\hbox{\scriptsize SLS}})\right]^2 \, .
\end{eqnarray}
In figure \ref{Fig:C_Theta_NSW} we show as a dashed curve the evaluation of
eq.\,(\ref{Eq:C_theta_NSW}) for $\Omega_{\hbox{\scriptsize tot}}=1.020$.
The solid curve shows the full result including the integrated
Sachs-Wolfe and Doppler contribution.
The grey band corresponds to the $1\sigma$ deviation
computed from 500 simulations.
Although the curve corresponding to the ordinary Sachs-Wolfe contribution
lies within the $1\sigma$ band of the full result,
the other two contributions cannot be neglected
as discussed in more detail below.

In our computations of the multipole moments $C_l$ for a given
primordial realization,
we use all eigenmodes with $\beta\le 155$.
Above this cut-off, we use in the mode summation up to $\beta=1501$
the conjecture (\ref{Eq:Conjecture})
which describes the mean asymptotic behaviour.
If only the modes below $\beta=155$ are used,
the multipoles can only be calculated up to $l\simeq 15$.
It is thus important for the calculation of $C(\vartheta)$
that the conjecture allows us to include higher multipoles.

Let us now come to the discussion of the CMB anisotropy
in the Poincar\'e dodecahedral space,
where, for simplicity, we assume that the dark energy is given by
a cosmological constant with
$\Omega_\Lambda := \Omega_{\hbox{\scriptsize tot}} -
\Omega_{\hbox{\scriptsize rad}} - \Omega_{\hbox{\scriptsize mat}}$.
We have calculated the CMB anisotropy for different values of the
cosmological parameters.
Here we present only the results for
$\Omega_{\hbox{\scriptsize bar}}=0.046$,
$\Omega_{\hbox{\scriptsize mat}}=0.28$ and $h=0.70$.

\begin{figure}[htb]
\begin{center}
\vspace*{-60pt}
\hspace*{-130pt}
\begin{minipage}{6cm}
\includegraphics[width=9.0cm]{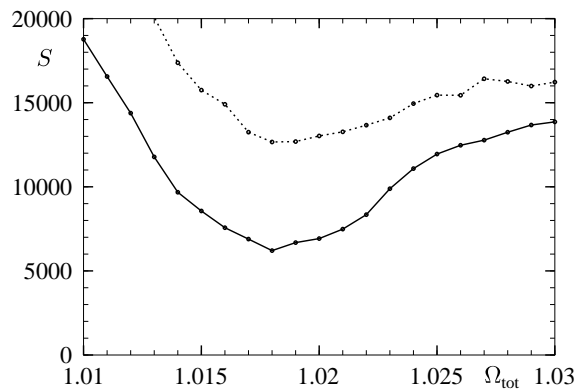}
\end{minipage}
\vspace*{-40pt}
\end{center}
\caption{\label{Fig:S_statistic_median}
The values of the $S(\rho)$ statistic (\ref{Eq:S-Statistik})
for the median values of $C(\vartheta)$ for the dodecahedral space
for $\rho=60^\circ$ (full curve) and $\rho=20^\circ$ (dotted curve)
in dependence on $\Omega_{\hbox{\scriptsize tot}}$
($h=0.70$,
$\Omega_{\hbox{\scriptsize mat}}=0.28$ and
$\Omega_{\hbox{\scriptsize bar}}=0.046$).
}
\end{figure}

To quantify the observed surprisingly low CMB anisotropy
\cite{Bennett_et_al_2003} on large scales, the $S$ statistic
\begin{equation}
\label{Eq:S-Statistik}
S(\rho) \; = \;
\int_{-1}^{\cos\rho} \big| C(\vartheta) \big|^2 \; d\cos\vartheta
\end{equation}
is discussed for the first year WMAP data in \cite{Spergel_et_al_2003}
for $\rho=60^\circ$,
and it is found that only $0.3\%$ of the simulations based on the
concordance model ri have lower values of $S(60^\circ)$
than the observed value $S(60^\circ)=1644$.
Somewhat higher values of $S(60^\circ)$ are obtained using other
statistical methods and other sky masks in \cite{Efstathiou_2004},
but they are nevertheless surprisingly low.
In figure \ref{Fig:S_statistic_median} we show the values of
$S(60^\circ)$ (full curve) and $S(20^\circ)$ (dotted curve)
using the median values of $C(\vartheta)$
for the dodecahedral space ${\cal D}$ in dependence on
$\Omega_{\hbox{\scriptsize tot}}$.
One observes that the models with 
$\Omega_{\hbox{\scriptsize tot}} = 1.017\dots 1.020$
give the lowest values for the large scale anisotropy.

Here and in the following calculations,
the overall amplitude of the CMB anisotropy,
i.\,e.\ the normalization constant $\alpha$ in (\ref{Eq:Harrison_Zeldovich}),
is fitted to the $C_l$ values of the first year WMAP data in the range 
$l\in[20,45]$.
(Thus, the values cannot be directly compared to the ones in
\cite{Luminet_Weeks_Riazuelo_Lehoucq_Uzan_2003},
where the amplitudes are scaled such that they match the $C_4$ value
of WMAP exactly.)

\begin{figure}[htb]
\begin{center}
\vspace*{-70pt}\hspace*{-80pt}\begin{minipage}{14cm} 
\begin{minipage}{6cm}
\includegraphics[width=9.0cm]{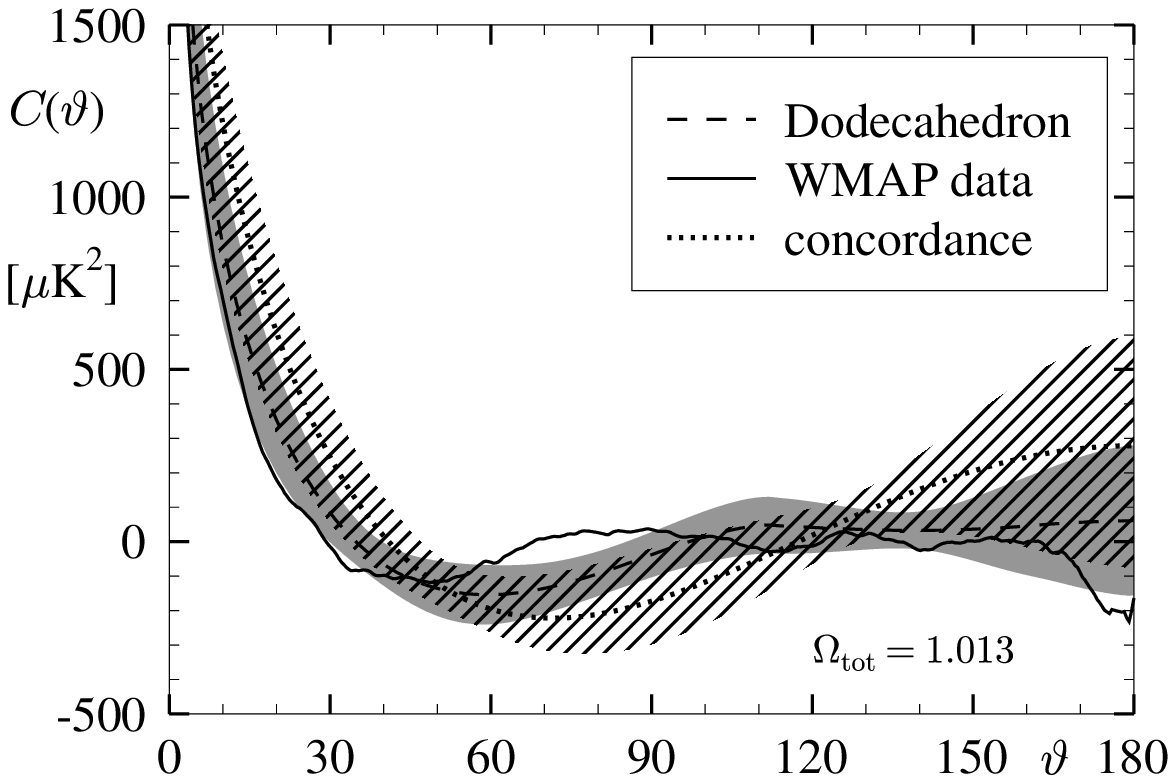}
\put(-160,150){a)}
\end{minipage}
\begin{minipage}{6cm}
\hspace*{40pt}\includegraphics[width=9.0cm]{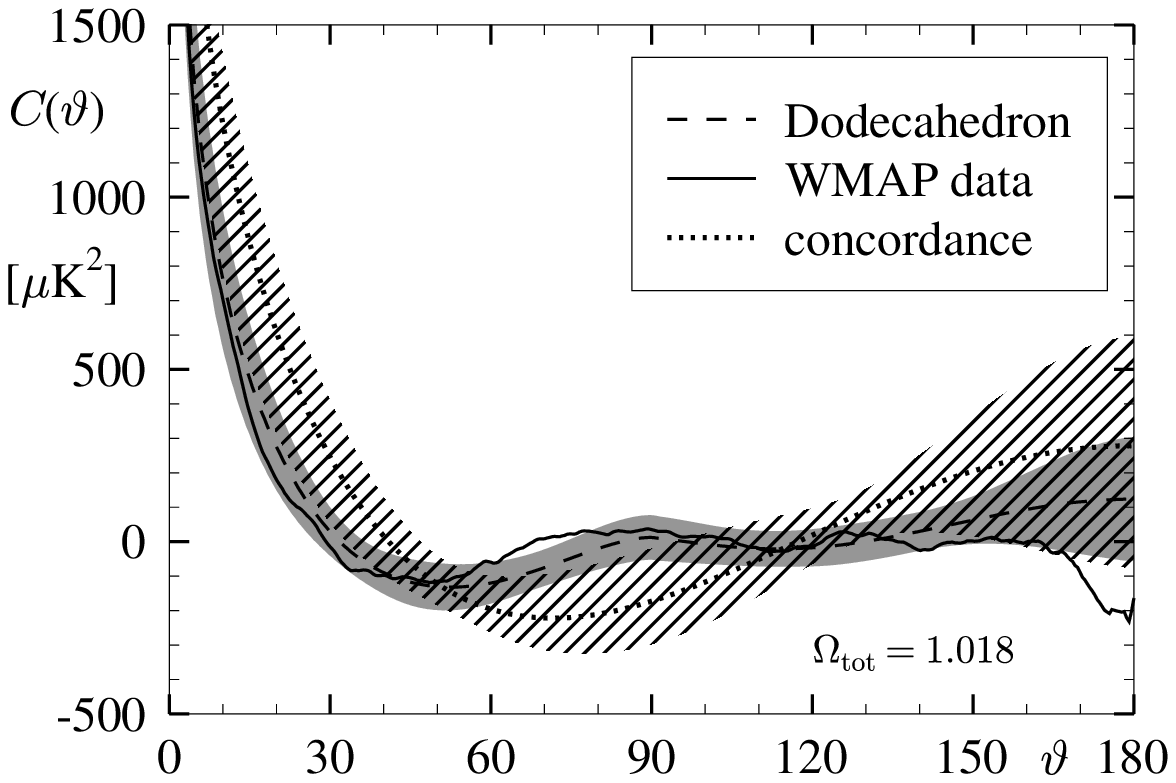}
\put(-160,150){b)}
\end{minipage}
\end{minipage}
\hspace*{-80pt}\begin{minipage}{14cm} 
\vspace*{-100pt}\begin{minipage}{6cm}
\includegraphics[width=9.0cm]{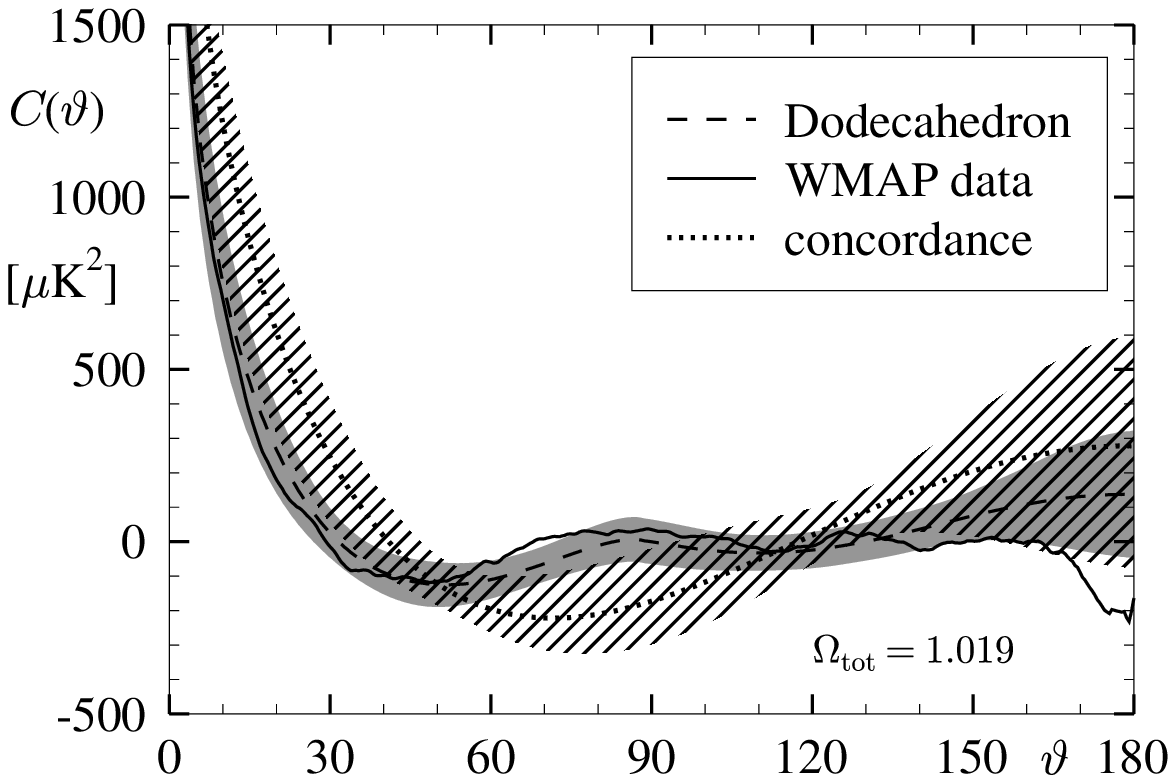}
\put(-160,150){c)}
\end{minipage}
\begin{minipage}{6cm}
\hspace*{40pt}\includegraphics[width=9.0cm]{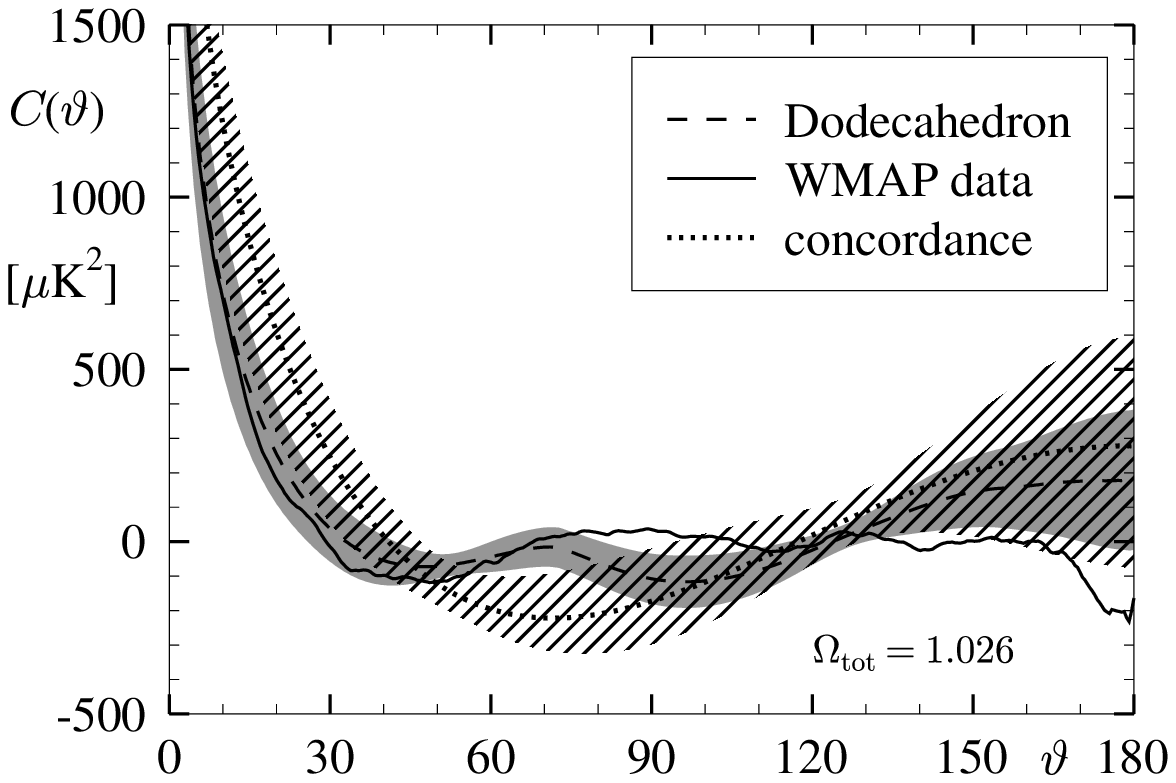}
\put(-160,150){d)}
\end{minipage}
\end{minipage}
\vspace*{-40pt}
\end{center}
\caption{\label{Fig:C_theta}
The temperature correlation function $C(\vartheta)$
for the dodecahedral topology for four values
of $\Omega_{\hbox{\scriptsize tot}}$ for $h=0.70$,
$\Omega_{\hbox{\scriptsize mat}}=0.28$ and
$\Omega_{\hbox{\scriptsize bar}}=0.046$,
the WMAP first year observation and the pl-concordance model
(see description in the text).
}
\end{figure}

In figure \ref{Fig:C_theta} the temperature correlation function
$C(\vartheta)$ is shown for four values of $\Omega_{\hbox{\scriptsize tot}}$.
The WMAP curve obtained from the LAMBDA home page
{\it http://lambda.gsfc.nasa.gov} is shown as a full curve.
The mean value of $C(\vartheta)$ for the pl-concordance model
from the same home page is displayed as dotted curve.
The $1\sigma$ deviation computed from 3000 HEALPix
\cite{Gorski_Hivon_Wandelt_1999} simulations is shown as a shaded region.
Finally, the mean value of $C(\vartheta)$ for the
dodecahedral topology is shown as a dashed curved and the
corresponding  $1\sigma$ deviation is the grey region
obtained from 500 simulations.
In figure \ref{Fig:C_theta}a), the value preferred in
\cite{Luminet_Weeks_Riazuelo_Lehoucq_Uzan_2003},
i.\,e.\ $\Omega_{\hbox{\scriptsize tot}}=1.013$, is presented.
One observes a general low anisotropy on large scales,
however, a more than $1\sigma$ deviation from the WMAP values
occurs in the range $\vartheta\in[60^\circ,90^\circ]$.
The WMAP data are much better described by models with 
$\Omega_{\hbox{\scriptsize tot}}=1.018$ and
$\Omega_{\hbox{\scriptsize tot}}=1.019$
(figures \ref{Fig:C_theta}b) and \ref{Fig:C_theta}c))
as is also revealed by the better values for the $S$ statistic
(compare figure \ref{Fig:S_statistic_median}).
Only on the largest scales $\vartheta>170^\circ$,
the models in figures \ref{Fig:C_theta}b) and \ref{Fig:C_theta}c)
do not reproduce the large negative correlation obtained from the WMAP data.
But without this blemish, the dodecahedral space describes
in the considered $\Omega_{\hbox{\scriptsize tot}}$ range the observations
down to the smallest scales,
since the WMAP curve lies completely within the grey $1\sigma$ region.
Finally, figure \ref{Fig:C_theta}d) shows a model
with $\Omega_{\hbox{\scriptsize tot}}=1.026$,
where the $S$ statistic already deteriorates and
consequently the agreement with the WMAP curve is also worse.

\begin{figure}[htb]
\begin{center}
\vspace*{-70pt}\hspace*{-80pt}\begin{minipage}{14cm} 
\begin{minipage}{6cm}
\includegraphics[width=9.0cm]{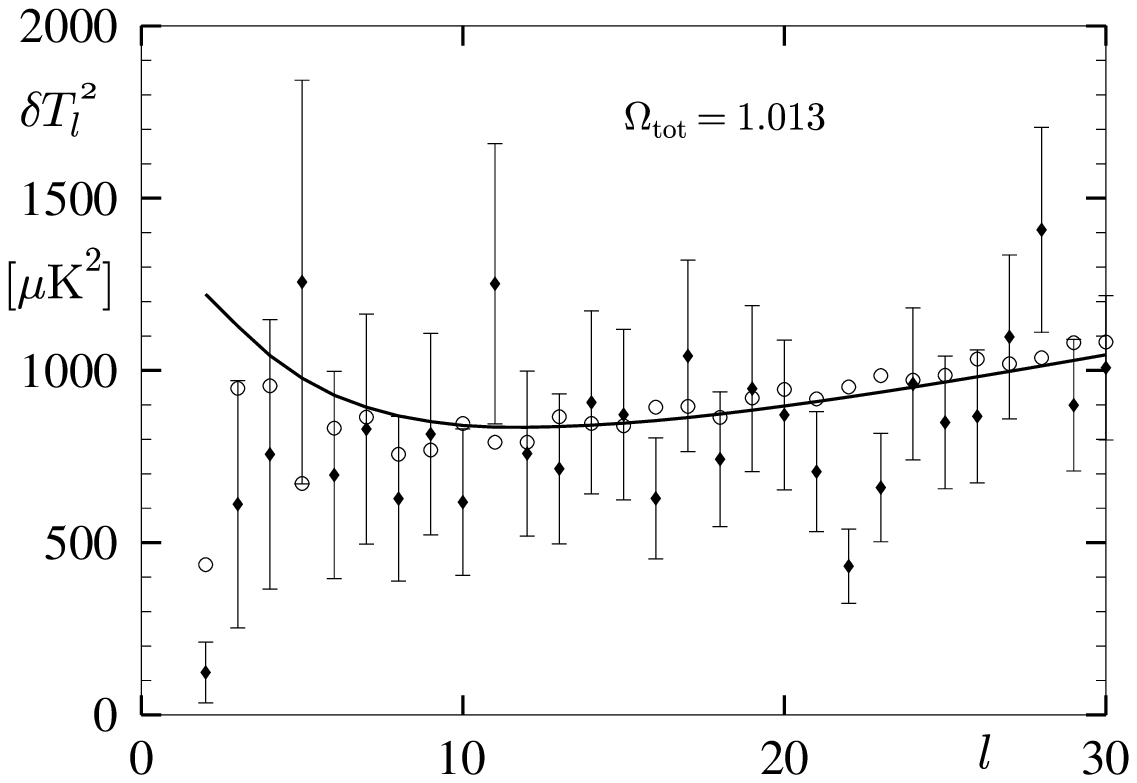}
\put(-178,150){a)}
\end{minipage}
\begin{minipage}{6cm}
\hspace*{40pt}\includegraphics[width=9.0cm]{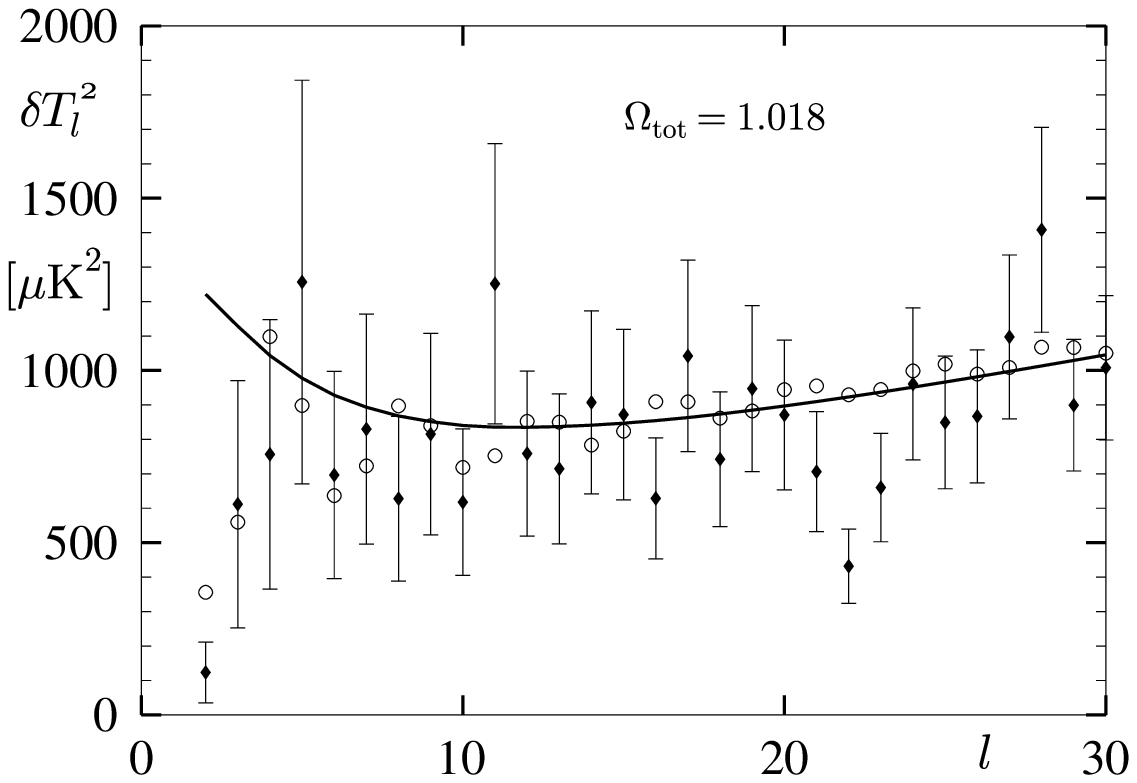}
\put(-178,150){b)}
\end{minipage}
\end{minipage}
\hspace*{-80pt}\begin{minipage}{14cm} 
\vspace*{-100pt}\begin{minipage}{6cm}
\includegraphics[width=9.0cm]{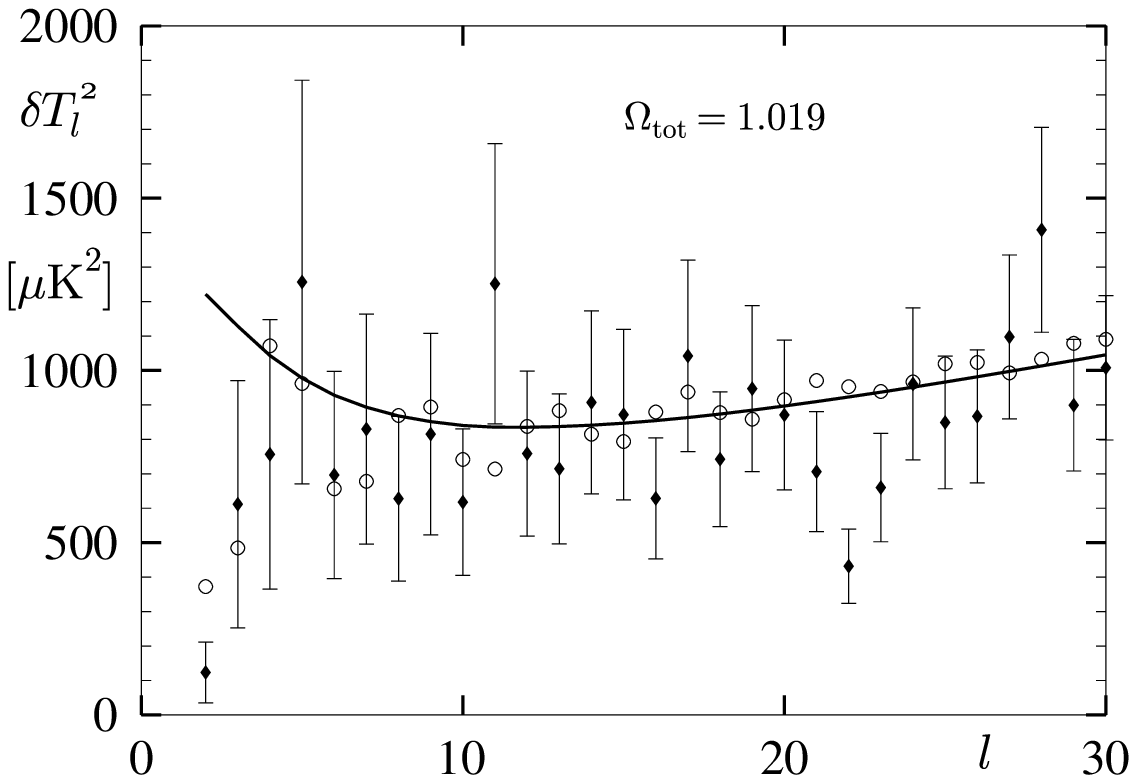}
\put(-178,150){c)}
\end{minipage}
\begin{minipage}{6cm}
\hspace*{40pt}\includegraphics[width=9.0cm]{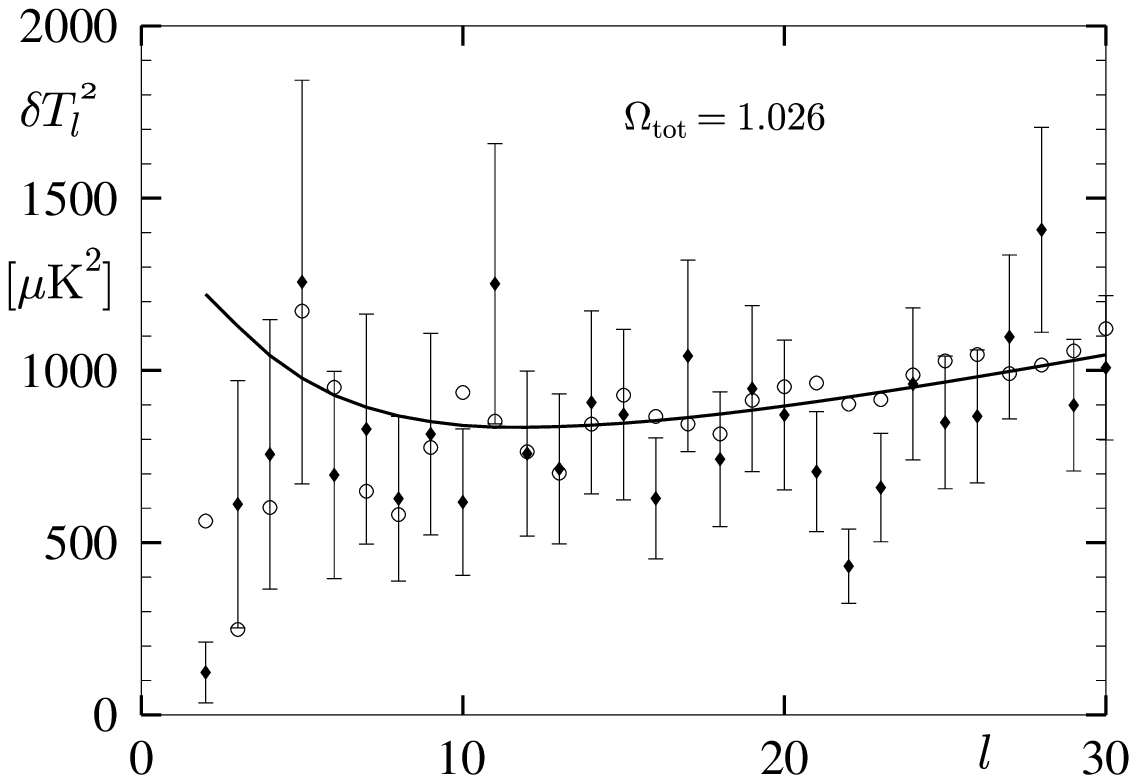}
\put(-178,150){d)}
\end{minipage}
\end{minipage}
\vspace*{-40pt}
\end{center}
\caption{\label{Fig:C_l}
The angular power spectrum $\delta T_l^2$ for the
dodecahedral topology (circles) for four values
of $\Omega_{\hbox{\scriptsize tot}}$ for $h=0.70$,
$\Omega_{\hbox{\scriptsize mat}}=0.28$ and
$\Omega_{\hbox{\scriptsize bar}}=0.046$.
The WMAP data are shown as diamonds with error bars, and
the pl-concordance model is represented by the full curve.
}
\end{figure}

\begin{figure}[tbh]
\begin{center}
\vspace*{-90pt}\hspace*{-90pt}\begin{minipage}{7cm}
\includegraphics[width=11.0cm]{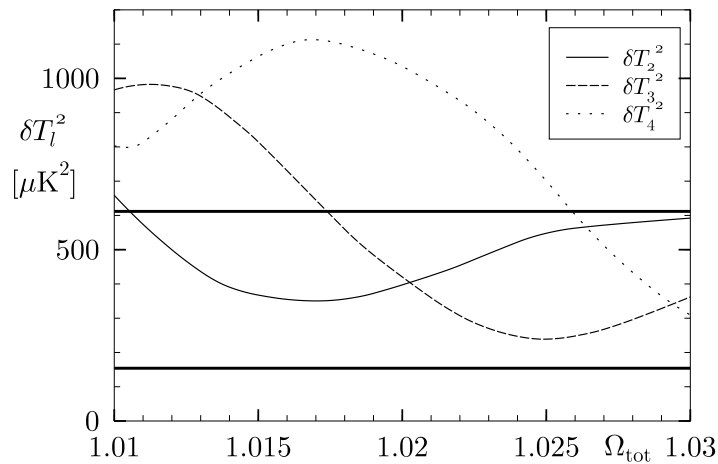}
\end{minipage}
\vspace*{-45pt}
\end{center}
\caption{\label{Fig:C_l_Omega_tot}
The mean values of $\delta T_2^2$ (solid curve), $\delta T_3^2$ (dashed curve)
and $\delta T_4^2$ (dotted curve) are shown for the
dodecahedral space ${\cal D}$ as a function
of $\Omega_{\hbox{\scriptsize tot}}$.
The lower horizontal line indicates the $\delta T_2^2$ value of WMAP
and the upper one the corresponding value of $\delta T_3^2$.
}
\end{figure}

In figures \ref{Fig:C_l}a) to \ref{Fig:C_l}d), the angular power spectrum
$\delta T_l^2= l(l+1) C_l /(2\pi)$ is shown for the same values of
$\Omega_{\hbox{\scriptsize tot}}$ as in figure \ref{Fig:C_theta}.
The WMAP data are represented by diamonds together with the
$1\sigma$ error bars which do not include the cosmic variance.
The mean values of $\delta T_l^2$ of the pl-concordance model (full curve)
do not show the suppression of power at low values of $l$,
but in contrast, increase there.
The mean values of $\delta T_l^2$ for the dodecahedral space exhibit
the suppression of power as observed by WMAP.
A good agreement is observed exactly for those values of
$\Omega_{\hbox{\scriptsize tot}}$ favored by the $S$ statistic
and the correlation function $C(\vartheta)$.
This behaviour is also revealed by figure \ref {Fig:C_l_Omega_tot}
which shows the dependence on $\Omega_{\hbox{\scriptsize tot}}$
of the mean values of $\delta T_2^2$, $\delta T_3^2$ and $\delta T_4^2$
for the dodecahedral space ${\cal D}$.
The lowest values of the quadrupole $\delta T_2^2$ are obtained near
$\Omega_{\hbox{\scriptsize tot}} = 1.017$.
Furthermore, at $\Omega_{\hbox{\scriptsize tot}} = 1.017$
the next multipole $\delta T_3^2$ matches the value obtained
by WMAP almost perfectly.

\begin{figure}[htb]
\begin{center}
\vspace*{-60pt}
\hspace*{-130pt}\begin{minipage}{14cm} 
\begin{minipage}{6cm}
\includegraphics[width=9.0cm]{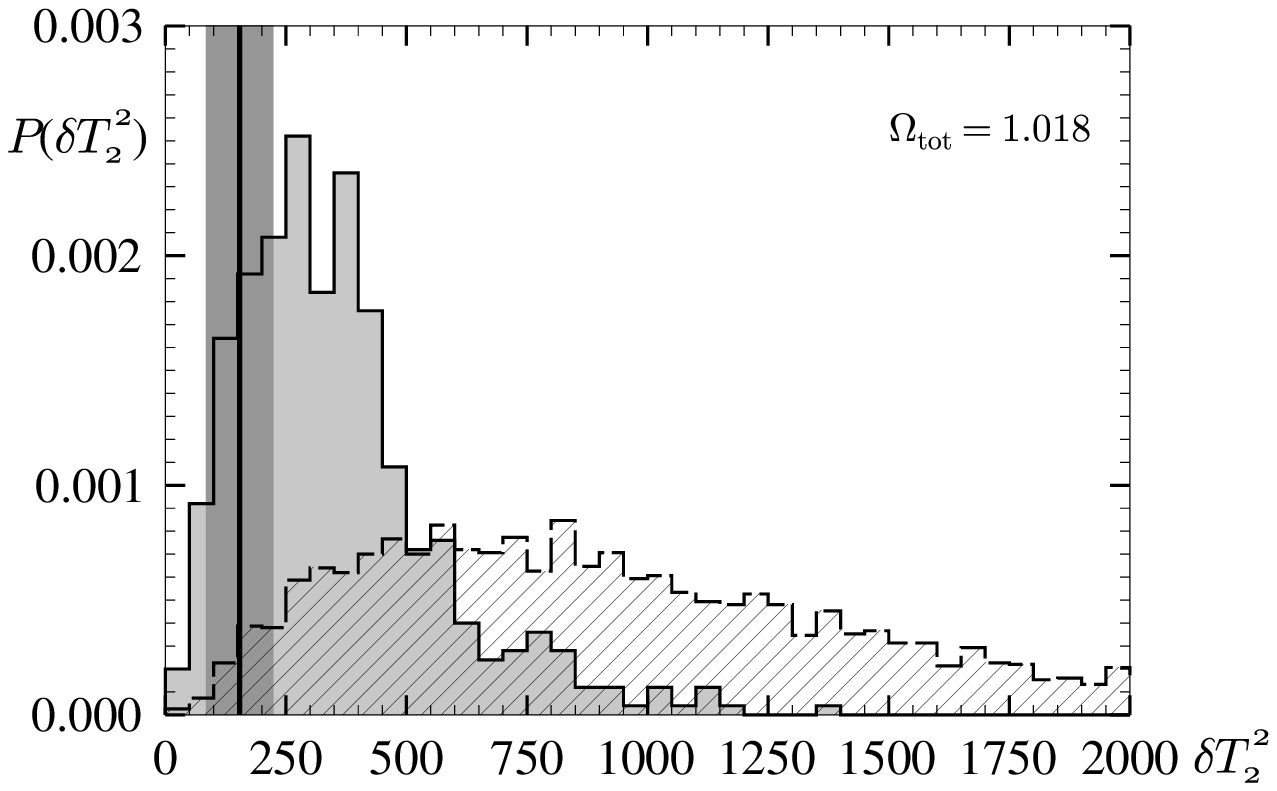}
\end{minipage}
\begin{minipage}{6cm}
\hspace*{70pt}\includegraphics[width=9.0cm]{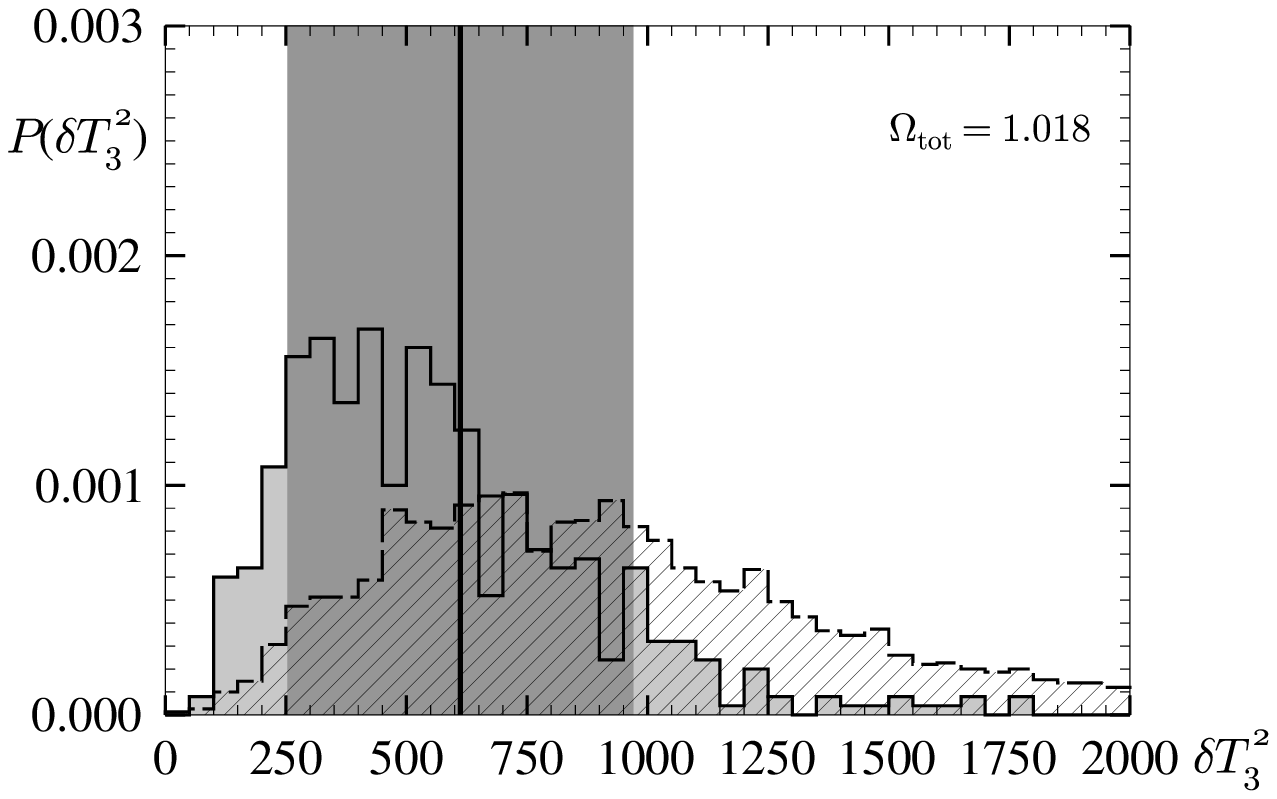}
\end{minipage}
\end{minipage}
\vspace*{-40pt}
\end{center}
\caption{\label{Fig:C_23}
The probability distribution of $\delta T_2^2$ and $\delta T_3^2$
in $[\mu\hbox{K}^2]$ for $\Omega_{\hbox{\scriptsize tot}}=1.018$, $h=0.70$,
$\Omega_{\hbox{\scriptsize mat}}=0.28$ and
$\Omega_{\hbox{\scriptsize bar}}=0.046$
for the dodecahedral space (light grey histogram).
The wide distribution of the pl-concordance model is shown as the
shaded histogram.
The WMAP values together with their $1\sigma$ error are indicated
by the vertical line and the vertical grey band, respectively.
}
\end{figure}

The probability distribution of $\delta T_2^2$ and $\delta T_3^2$ obtained
from sky simulations is for the pl-concordance model much broader than for
the dodecahedral space.
To emphasize this fact,
figure \ref{Fig:C_23} displays the distribution of $\delta T_2^2$ and
$\delta T_3^2$ for $\Omega_{\hbox{\scriptsize tot}}=1.018$.
The corresponding WMAP values are indicated by the vertical lines
together with the $1\sigma$ error (vertical dark grey band).
The shaded histogram shows the wide distribution of 3000 HEALPix simulations
for the pl-concordance model.
In contrast, a much more pronounced peak is revealed by the corresponding
distribution of the dodecahedral space (light grey histogram).
Furthermore, the peaks for $\delta T_2^2$ and $\delta T_3^2$
lie both within the $1\sigma$ error of the WMAP observation.

\section{The circles-in-the-sky signature}

\label{circles-in-the-sky-signature}

A multi-connected space can reveal itself by the so-called
circles-in-the-sky signature
\cite{Cornish_Spergel_Starkman_1998b}.
Consider an observer located at a given point $\vec x$
who measures the fluctuations of the CMB.
The contribution to the CMB fluctuations due to the
ordinary Sachs-Wolfe (SW) effect arises on a sphere around this observer
having the radius $\tau_{\hbox{\scriptsize SLS}}$,
i.\,e.\ at the so-called surface of last scattering (SLS)
w.\,r.\,t.\ this observer.
In the universal cover there are copies of this observer
which have their own SLS's which are centered at the copies
of the observer.
Those copies whose distance to the observer at $\vec x$
are smaller than $2 \tau_{\hbox{\scriptsize SLS}}$ possess SLS's
which intersect the SLS of the observer sitting at $\vec x$.
Since the intersection of two spheres is a circle,
both the observer and its copy observe along this circle
the same temperature fluctuations due to the ordinary Sachs-Wolfe effect.
Since the observer and its copy have to be identified,
there are two matching  circles having the same radius on the sky
and having an identical ordinary Sachs-Wolfe contribution.
The integrated Sachs-Wolfe (ISW) contribution arises on the photon path
from the SLS to the observer, see eq.\,(\ref{Eq:Sachs_Wolfe_tight_coupling}).
Since the two paths are not identified,
their contribution is different on the matched circles.
The two Doppler contributions are also different,
since their velocity projections are different.

\begin{figure}[tbh]
\begin{center}
\vspace*{-60pt}
\hspace*{-130pt}\begin{minipage}{14cm} 
\begin{minipage}{6cm}
\includegraphics[width=9.0cm]{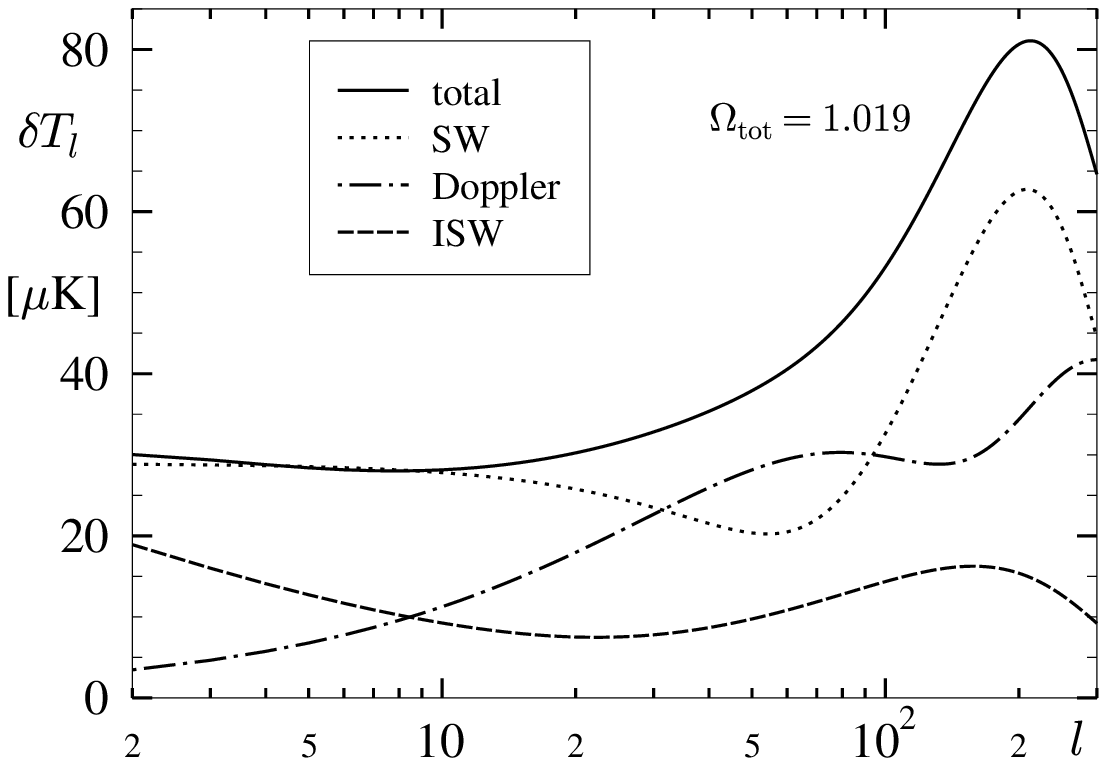}
\put(-180,150){a)}
\end{minipage}
\begin{minipage}{6cm}
\hspace*{70pt}\includegraphics[width=9.0cm]{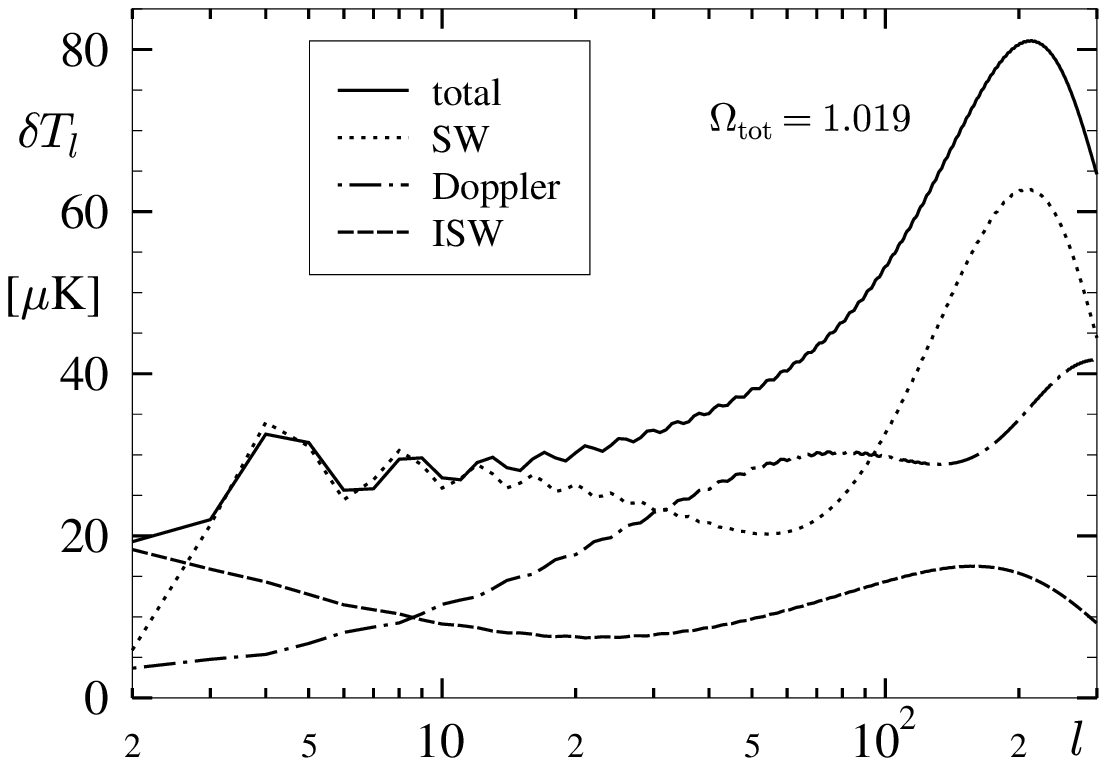}
\put(-180,150){b)}
\end{minipage}
\end{minipage}
\vspace*{-40pt}
\end{center}
\caption{\label{Fig:delta_T_NSW_ISW}
The three contributions to $\delta T_l$ are shown
for the simply connected ${\cal S}^3$ (left) and
the dodecahedral space ${\cal D}$ (right)
calculated in the tight-coupling approximation for
$\Omega_{\hbox{\scriptsize tot}}=1.019$, $h=0.70$,
$\Omega_{\hbox{\scriptsize mat}}=0.28$ and
$\Omega_{\hbox{\scriptsize bar}}=0.046$.
}
\end{figure}

In the case of a flat toroidal universe, these different contributions 
to the temperature anisotropy are investigated in
\cite{Riazuelo_Uzan_Lehoucq_Weeks_2004} with respect to the
circles-in-the-sky signature, and it is found
that the Doppler and the integrated Sachs-Wolfe contributions
degrade the signature.
That these two contributions play an important r\^ole also in the case of
the dodecahedral topology, is demonstrated in figure \ref{Fig:delta_T_NSW_ISW}.
It shows the three competing contributions as computed from
eq.\,(\ref{Eq:Sachs_Wolfe_tight_coupling}) for 
the simply connected ${\cal S}^3$ (figure \ref{Fig:delta_T_NSW_ISW}a)) and for
the dodecahedral space ${\cal D}$ (figure \ref{Fig:delta_T_NSW_ISW}b)).
Here we use the conjecture (\ref{Eq:Conjecture}) up to $\beta=4001$.
For the SW contribution, see eq.\,(\ref{Eq:multipol_NSW}).
The most obvious difference between ${\cal D}$ and ${\cal S}^3$ is the
suppression of power for small multipoles $\delta T_l$ in the case
of ${\cal D}$ in comparison with ${\cal S}^3$.
In the case of the dodecahedral space ${\cal D}$,
the loss of power is so severe that for the quadrupole,
the ISW dominates over the ordinary Sachs-Wolfe contribution.
A further distinction between ${\cal S}^3$ and ${\cal D}$ is
that $\delta T_l$ oscillates as a function of $l$ in the case
of the multi-connected space even for the mean values of $\delta T_l$
which are considered here.
With the exception of the smallest multipoles,
the ordinary Sachs-Wolfe contribution dominates up to $l\lesssim 30$.
However, even in this range the Doppler and the ISW contribution
provide a significant fraction of the total temperature fluctuation.
The ISW is most important for $l\lesssim 10$,
whereas the Doppler contribution increases for $l\gtrsim 10$
to even larger values than the ISW,
such that both contributions give together a large perturbation to the
pure circles-in-the-sky signature.

\begin{figure}[tbh]
\begin{center}
\vspace*{-70pt}\hspace*{-80pt}\begin{minipage}{14cm} 
\begin{minipage}{6cm}
\includegraphics[width=9.0cm]{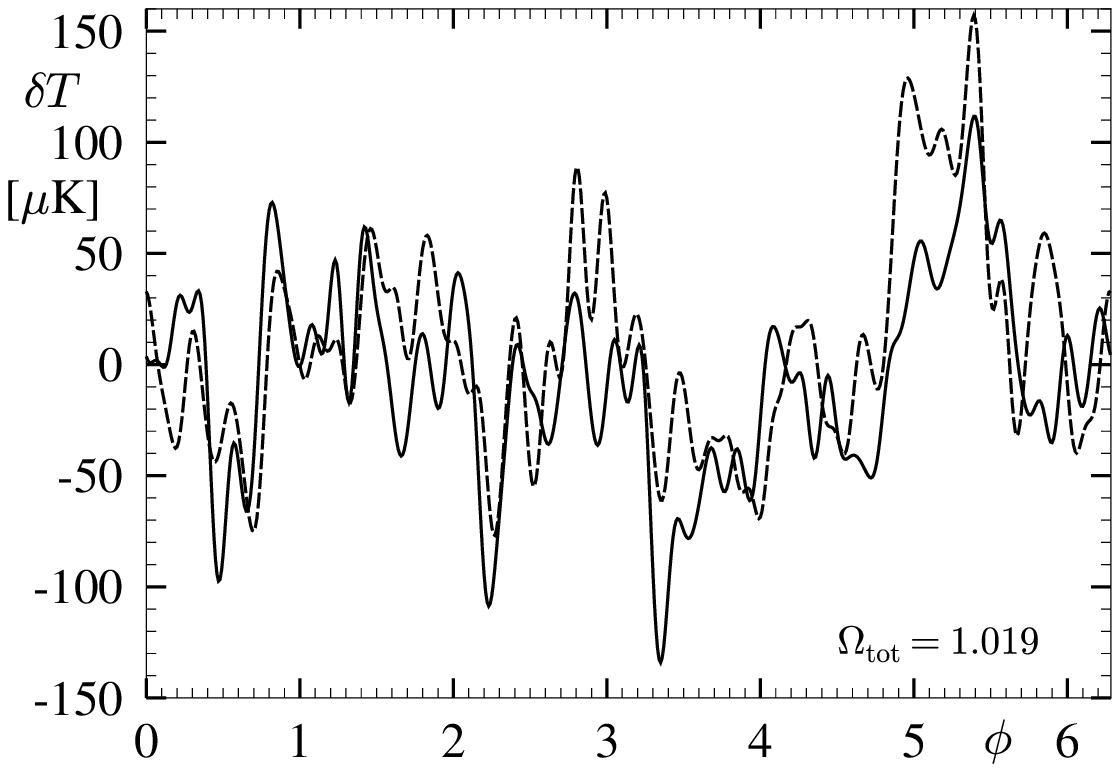}
\put(-180,150){a)}
\end{minipage}
\begin{minipage}{6cm}
\hspace*{40pt}\includegraphics[width=9.0cm]{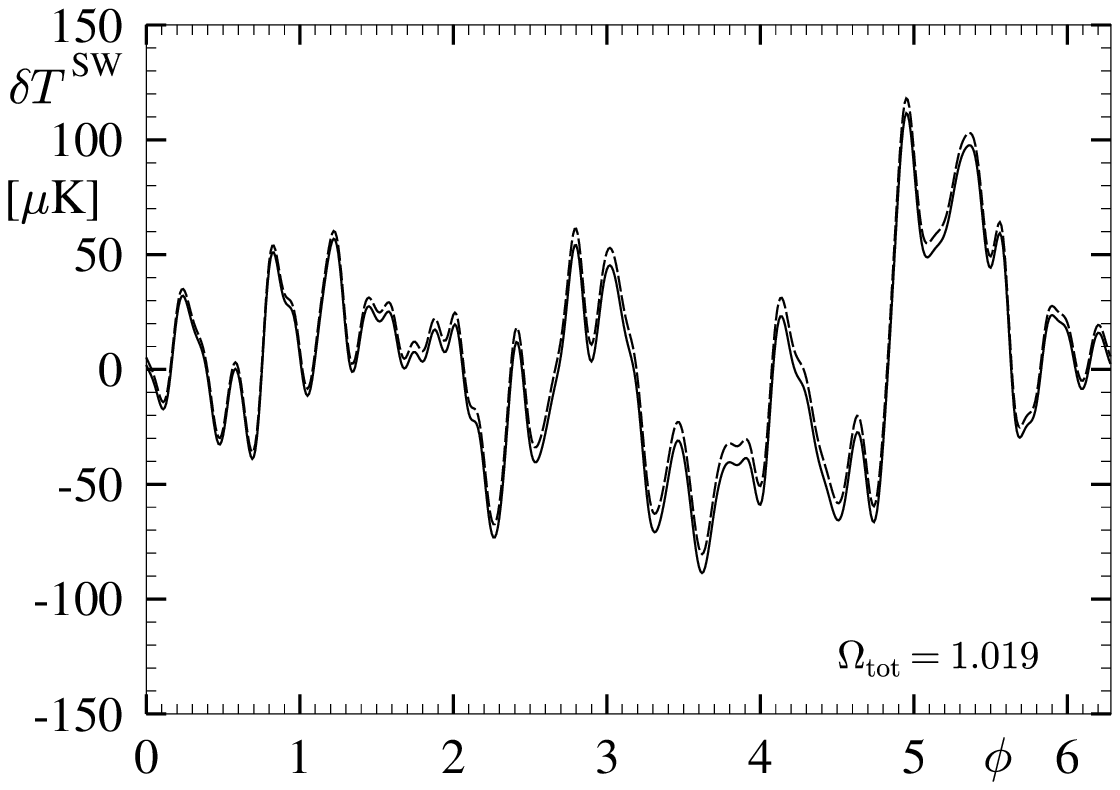}
\put(-180,150){b)}
\end{minipage}
\end{minipage}
\hspace*{-76pt}\begin{minipage}{14cm} 
\vspace*{-100pt}\begin{minipage}{6cm}
\includegraphics[width=9.0cm]{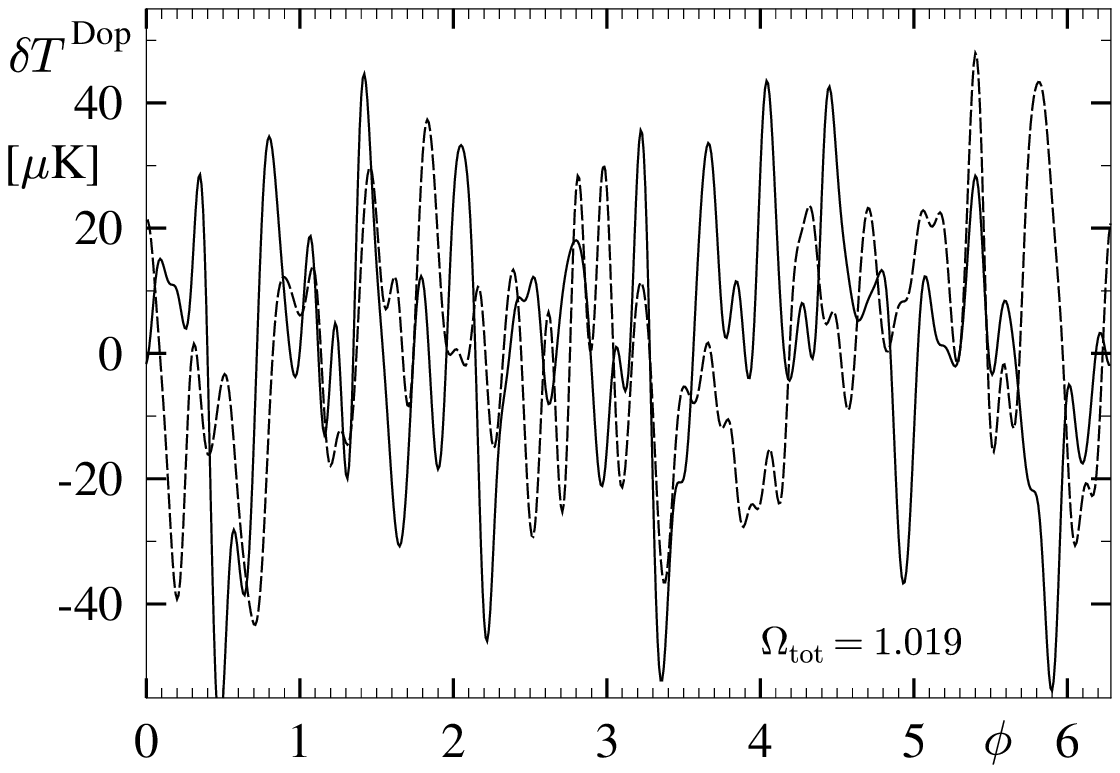}
\put(-180,150){c)}
\end{minipage}
\begin{minipage}{6cm}
\hspace*{40pt}\includegraphics[width=9.0cm]{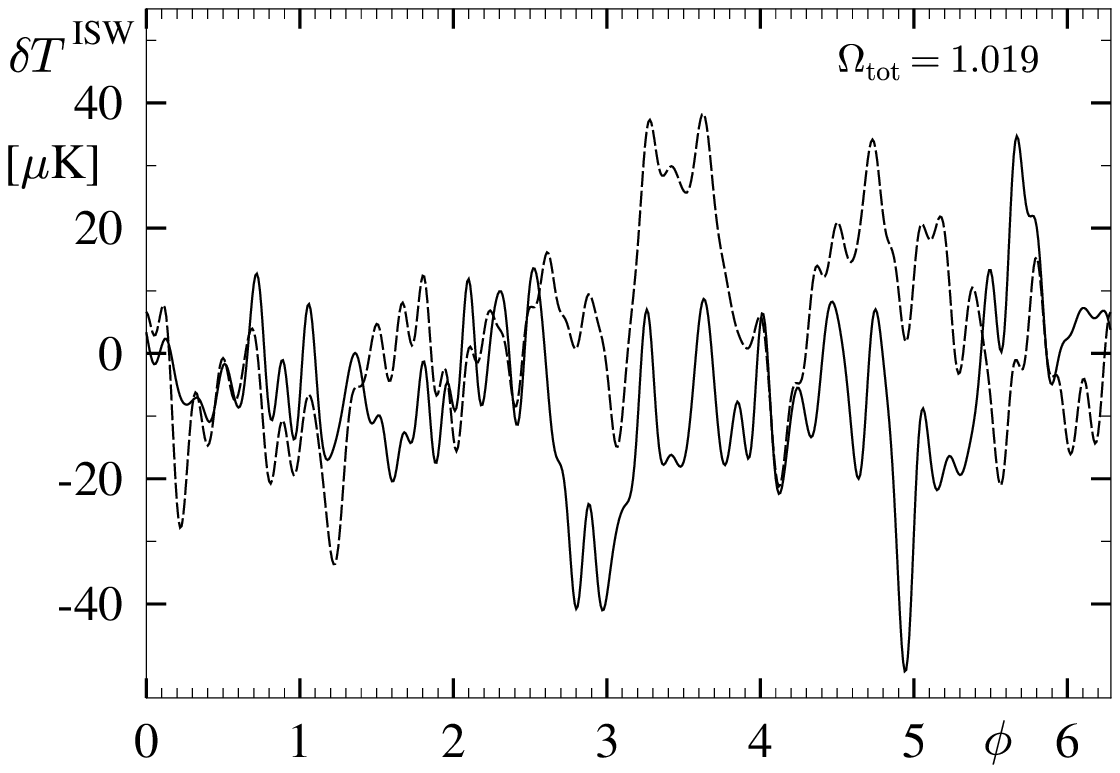}
\put(-180,150){d)}
\end{minipage}
\end{minipage}
\vspace*{-40pt}
\end{center}
\caption{\label{Fig:delta_T_NSW_ISW_phi}
The temperature fluctuation $\delta T$ is shown along two matched circles for
$\Omega_{\hbox{\scriptsize tot}}=1.019$, $h=0.70$,
$\Omega_{\hbox{\scriptsize mat}}=0.28$ and
$\Omega_{\hbox{\scriptsize bar}}=0.046$.
Panel a) displays $\delta T$ of the total temperature fluctuation
along the two circles,
obtained from the ordinary Sachs-Wolfe (SW), Doppler (Dop) and
integrated Sachs-Wolfe (ISW) contribution, which are shown
separately in panels b), c) and d), respectively.
}
\end{figure}

To emphasize this point, figure \ref{Fig:delta_T_NSW_ISW_phi} shows
the temperature fluctuation along two matched circles
parameterized by the angle $\phi$ defined in (\ref{Eq:S_measure}) below.
One observes in figure \ref{Fig:delta_T_NSW_ISW_phi}a)
that the total temperature fluctuation $\delta T$
is not matched perfectly well.
However, some rough similarities between both $\delta T$ curves are
nevertheless visible.
Whether such similarities occur sufficiently frequently by chance
in the simply connected ${\cal S}^3$ model,
such that the circles-in-the-sky signature is swamped, is discussed below.
The different contributions to $\delta T$ are separately shown
in figures \ref{Fig:delta_T_NSW_ISW_phi}b), \ref{Fig:delta_T_NSW_ISW_phi}c)
and \ref{Fig:delta_T_NSW_ISW_phi}d).
Even the ordinary Sachs-Wolfe contribution
(figure \ref{Fig:delta_T_NSW_ISW_phi}b)) does not provide a perfect match
which is due to the fact that the dipole contribution has been subtracted
from the sky map analogously as it is done on the observational side.
This destroys the perfect agreement due to the SW contribution.
Conversely, if one would know the topological structure of the Universe,
this would offer an opportunity to determine the dipole contribution
due to the primordial fluctuations
which is usually superseded by the Doppler shift due to our local motion.
The Doppler and the ISW contributions are completely different
for the two circles as seen in figures \ref{Fig:delta_T_NSW_ISW_phi}c) and
\ref{Fig:delta_T_NSW_ISW_phi}d), respectively, for reasons described above.

As just discussed, for cosmological models near
$\Omega_{\hbox{\scriptsize tot}}\simeq1.02$,
the ordinary Sachs-Wolfe contribution dominates for $l \lesssim 30$.
The value $l=30$ corresponds to a scale
$\theta \simeq \frac{180^\circ}l = 6^\circ$.
Thus the circles are blurred on scales below $\sim 6^\circ$.
On larger scales, the integrated Sachs-Wolfe and the Doppler contribution
lead to a modulation of matched circle structure.

For a quantitative search of matched circles in microwave sky maps,
the quantity
\begin{equation}
\label{Eq:S_measure}
\Sigma(\rho) \; := \;
\frac{\left< 2\, \delta T_a(\pm\phi)\, \delta T_b(\phi+\rho) \right>}
{\left< \delta T_a^2(\phi) +  \delta T_b^2(\phi) \right>}
\end{equation}
is introduced in \cite{Cornish_Spergel_Starkman_1998b}.
(In \cite{Cornish_Spergel_Starkman_1998b}, $\Sigma$ is called $S$,
a name we avoid in the following in order to not confuse the
reader with the previously discussed $S$ statistic.)
Here $\delta T_a(\phi)$ and $\delta T_b(\phi)$ are the temperature fluctuations
along two circles on the SLS with the same radius,
and $\left< \right> := \frac 1{2\pi}\int_0^{2\pi} d\phi$.
The angle $\rho$ describes the relative phase between the temperature
fluctuations on the two circles and
the plus/minus sign in the nominator allows
for orientable as well as for non-orientable manifolds.
The $\Sigma$ variable takes the value $\Sigma=+1$ in the case of a
perfect correlation and $\Sigma=-1$ for anticorrelated
temperature fluctuations.

\begin{figure}[tbh]
\begin{center}
\vspace*{-90pt}\hspace*{-90pt}\begin{minipage}{7cm}
\includegraphics[width=11.0cm]{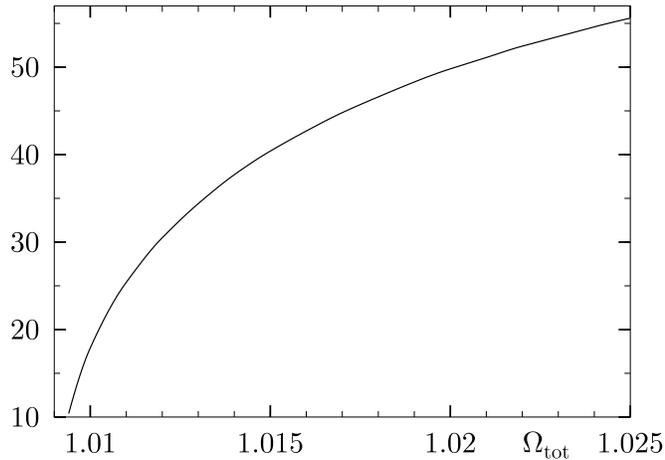}
\end{minipage}
\vspace*{-45pt}
\end{center}
\caption{\label{Fig:kreis_radius_omega}
The radius of the six matched circles is plotted as a function
of $\Omega_{\hbox{\scriptsize tot}}$.
}
\end{figure}

In \cite{Cornish_Spergel_Starkman_Komatsu_2003} the $\Sigma$ test
is applied to the first year data from WMAP and it is found
that $\Sigma_{\hbox{\scriptsize max}} = \max_\rho\{\Sigma(\rho)\}$
does not show any spikes which would reveal a matched circle pair.
The search is carried out for nearly back to back circles having a separation
larger than $170^\circ$ and a radius larger than $25^\circ$.
For $1.0091<\Omega_{\hbox{\scriptsize tot}}<1.0252$
(for $h=0.70$ and $\Omega_{\hbox{\scriptsize mat}}=0.28$)
the Poincar\'e dodecahedron has in total 6 pairs of circles
which are back to back, i.\,e.\ are separated by $180^\circ$.
The radius of these circles increases with increasing
$\Omega_{\hbox{\scriptsize tot}}$
as shown in figure \ref{Fig:kreis_radius_omega}.
Above $\Omega_{\hbox{\scriptsize tot}}=1.0252$ there are at first 16 pairs,
a number which increases further with increasing
$\Omega_{\hbox{\scriptsize tot}}$,
and below $\Omega_{\hbox{\scriptsize tot}}=1.0091$ there are none.
The conclusion in \cite{Cornish_Spergel_Starkman_Komatsu_2003} is thus
that the topology of the Poincar\'e dodecahedron is excluded.
However, in \cite{Roukema_et_al_2004} a hint towards six pairs of
circles having a radius of $11\pm 1^\circ$ is found
which would correspond to $\Omega_{\hbox{\scriptsize tot}}=1.0094$
for our choice of $h$ and $\Omega_{\hbox{\scriptsize mat}}$.

Now we would like to discuss whether the Doppler and ISW contributions
are significantly large enough in order to contaminate the SW contribution
such that the circles-in-the-sky signature is hidden.
We generate 500 sky maps for the dodecahedral space ${\cal D}$
using all modes up to $\beta\le 155$
and compute for each of the six pairs of matched circles the
value of $\Sigma_{\hbox{\scriptsize max}}$.
In this way we get 3000 values for $\Sigma_{\hbox{\scriptsize max}}$
belonging to circle pairs including the SW, Doppler and ISW contribution.
In addition, we obtain 3000 values for $\Sigma(\rho)$
where $\rho$ is not varied but instead is fixed to the value $\rho=36^\circ$
determined by the topology of ${\cal D}$.
Due to the Doppler and the ISW contribution, the values of
$\Sigma_{\hbox{\scriptsize max}}$ are usually larger than the correct
values $\Sigma(\rho)$ for the matched pair of circles having $\rho=36^\circ$.
This emphasizes that the contamination by the Doppler and the ISW
contribution is so large that the $\Sigma_{\hbox{\scriptsize max}}$
statistic does in general not find the correct value of $\rho$.
Furthermore, we compute also the corresponding values of
$\Sigma_{\hbox{\scriptsize max}}$ for 500 simulations for the
simply connected ${\cal S}^3$ universe with the same cut-off in $\beta$
and with the ring positions found in the dodecahedral space ${\cal D}$.
Of course, the $\Sigma_{\hbox{\scriptsize max}}$ values of
${\cal S}^3$ do not correspond to a matched circle pair
since ${\cal S}^3$ has none.

\begin{figure}[tbh]
\begin{center}
\vspace*{-80pt}
\hspace*{-130pt}\begin{minipage}{14cm} 
\begin{minipage}{6cm}
\includegraphics[width=9.0cm]{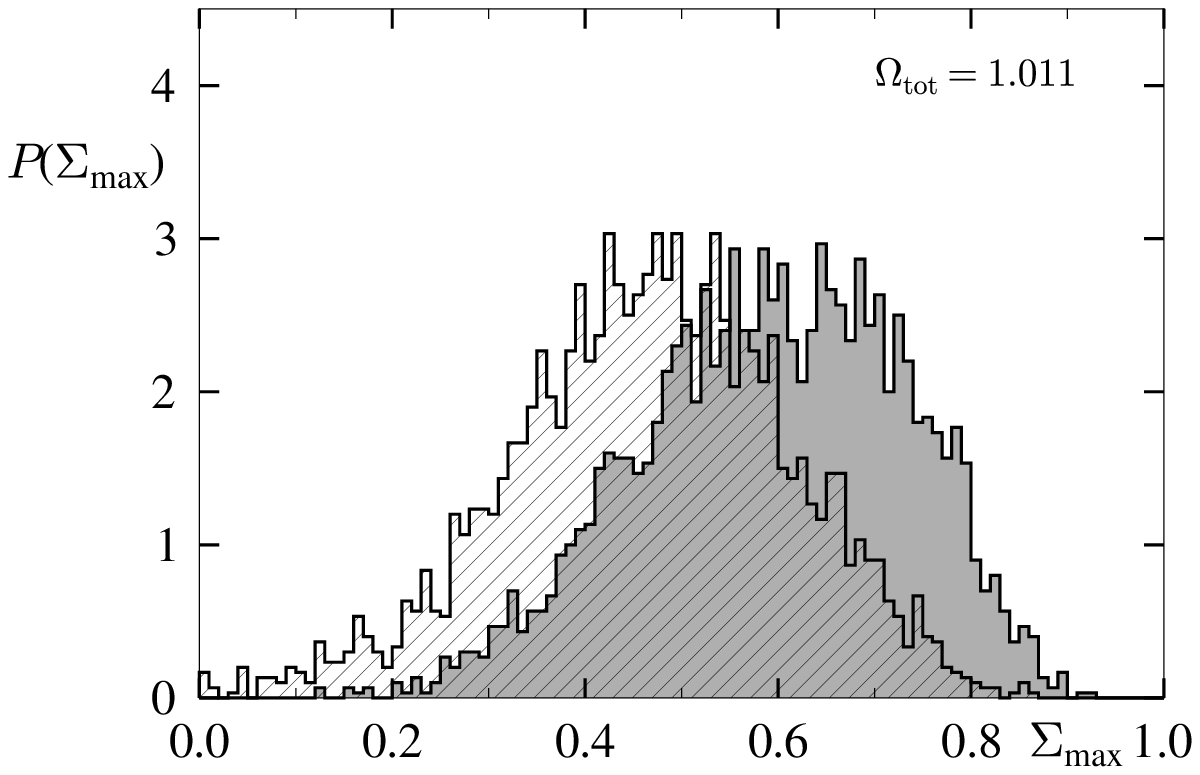}
\put(-170,130){a)}
\end{minipage}
\begin{minipage}{6cm}
\hspace*{70pt}\includegraphics[width=9.0cm]{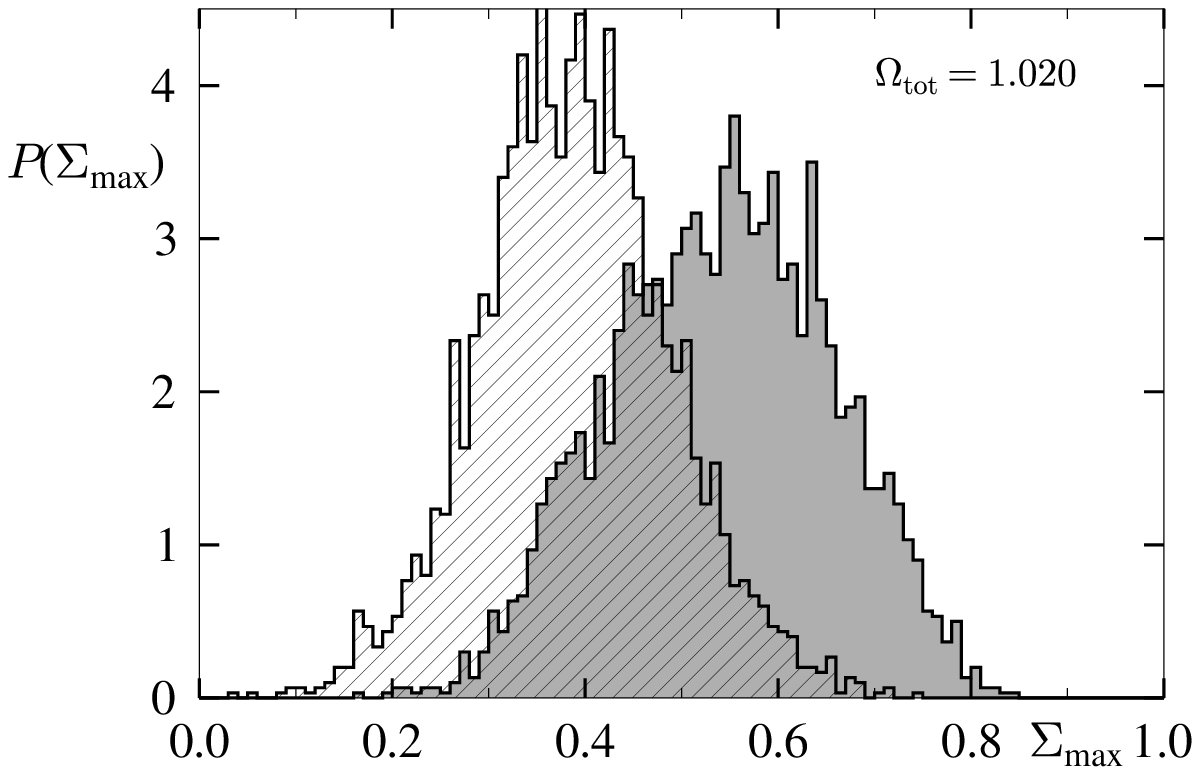}
\put(-170,130){b)}
\end{minipage}
\end{minipage}
\vspace*{-40pt}
\end{center}
\caption{\label{Fig:Sigma_distribution}
The probability distribution of $\Sigma_{\hbox{\scriptsize max}}$
for ${\cal S}^3$ (shaded histogram) and ${\cal D}$ (grey histogram)
for $\Omega_{\hbox{\scriptsize tot}}=1.011$ in panel a)
having a radius of $25.4^\circ$ and
for $\Omega_{\hbox{\scriptsize tot}}=1.020$ in panel b)
having a radius of $49.8^\circ$.
The other cosmological parameters are $h=0.70$,
$\Omega_{\hbox{\scriptsize mat}}=0.28$ and
$\Omega_{\hbox{\scriptsize bar}}=0.046$
as usual.
}
\end{figure}

\begin{figure}[tbh]
\begin{center}
\vspace*{-100pt}
\hspace*{-130pt}\begin{minipage}{14cm} 
\begin{minipage}{6cm}
\includegraphics[width=9.0cm]{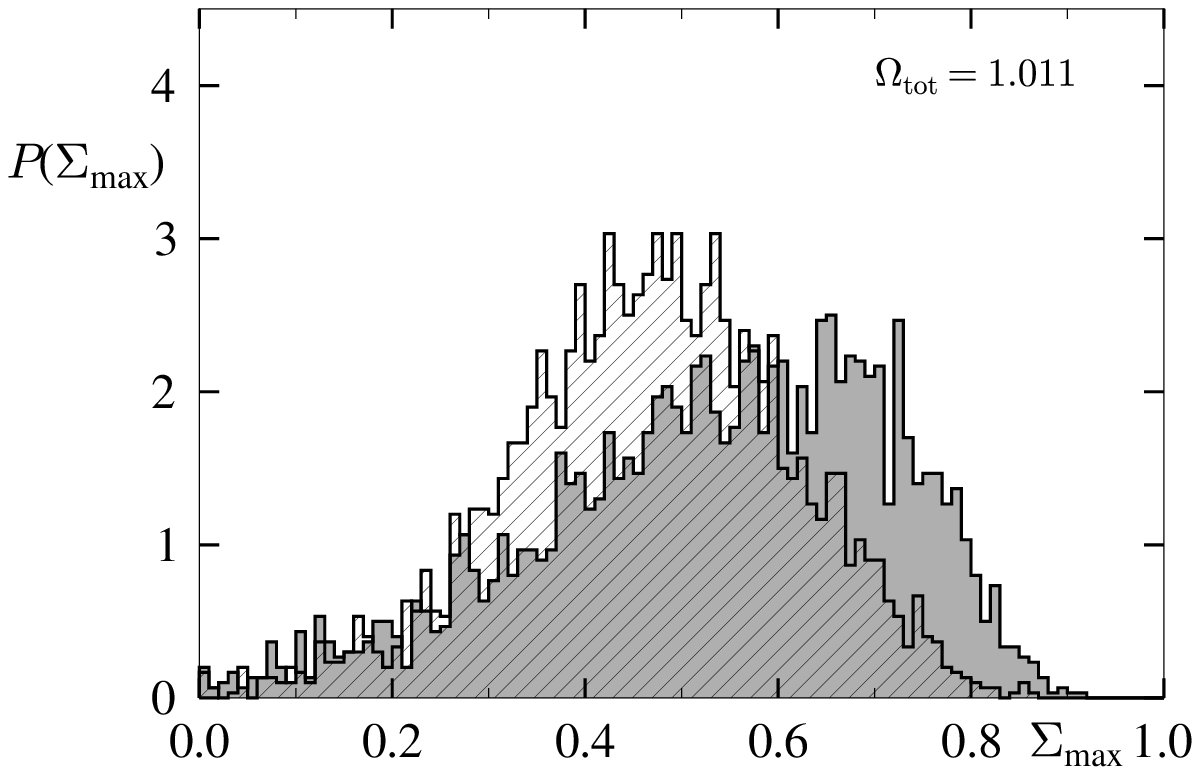}
\put(-170,130){a)}
\end{minipage}
\begin{minipage}{6cm}
\hspace*{70pt}\includegraphics[width=9.0cm]{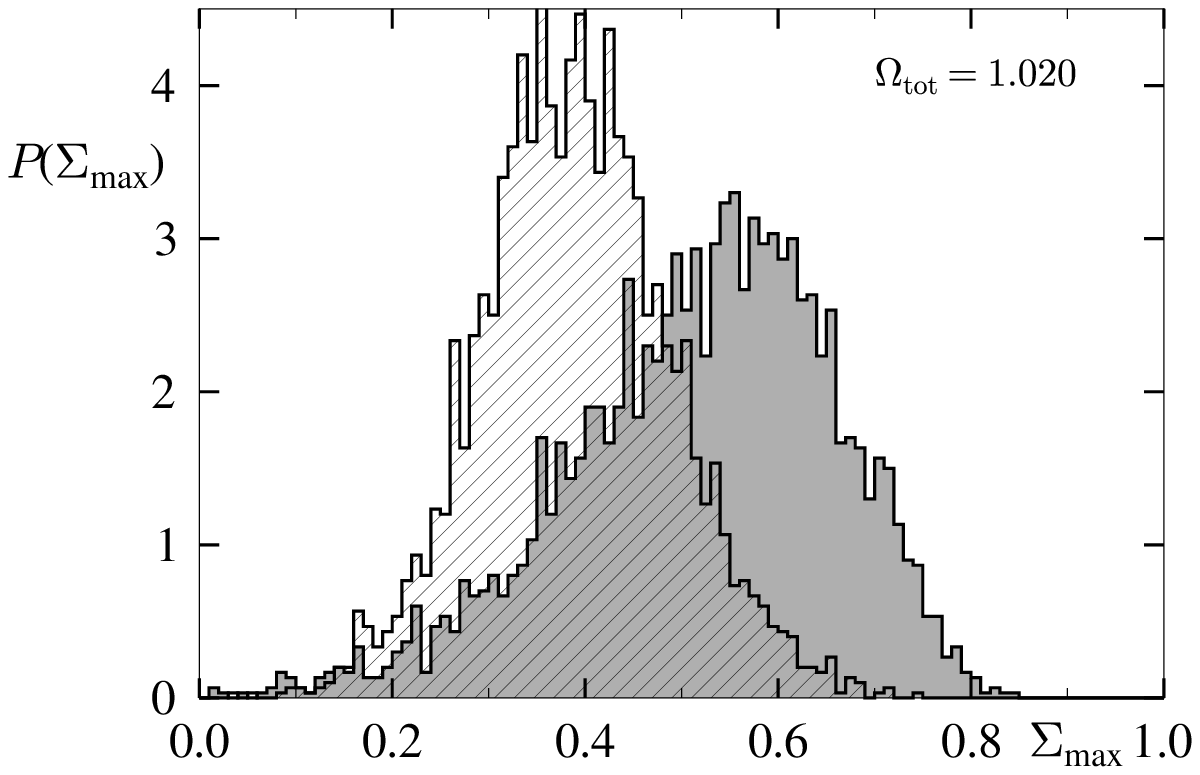}
\put(-170,130){b)}
\end{minipage}
\end{minipage}
\vspace*{-40pt}
\end{center}
\caption{\label{Fig:Sigma_distribution_no_rot}
The probability distribution of $\Sigma_{\hbox{\scriptsize max}}$
for ${\cal S}^3$ (shaded histogram) as in figure \ref{Fig:Sigma_distribution}
and of $\Sigma(\rho)$ for ${\cal D}$ with
$\rho=36^\circ$ as determined by the topology of ${\cal D}$ (grey histogram)
for $\Omega_{\hbox{\scriptsize tot}}=1.011$ in panel a)
having a radius of $25.4^\circ$ and
for $\Omega_{\hbox{\scriptsize tot}}=1.020$ in panel b)
having a radius of $49.8^\circ$.
The other cosmological parameters are 
as in figure \ref{Fig:Sigma_distribution}.
}
\end{figure}

In figure \ref{Fig:Sigma_distribution}, we compare the probability
distribution of $\Sigma_{\hbox{\scriptsize max}}$
for ${\cal S}^3$ and ${\cal D}$
for the two values $\Omega_{\hbox{\scriptsize tot}}=1.011$ and
$\Omega_{\hbox{\scriptsize tot}}=1.020$.
These values are chosen in order to demonstrate the effect due to the
size of the radii of the circles
which are $25.4^\circ$ and $49.8^\circ$ for
$\Omega_{\hbox{\scriptsize tot}}=1.011$ and
$\Omega_{\hbox{\scriptsize tot}}=1.020$, respectively. 
The probability distributions shift to smaller values of
$\Sigma_{\hbox{\scriptsize max}}$ with increasing radii.
In figure \ref{Fig:Sigma_distribution_no_rot}, we again show the
probability distribution of $\Sigma_{\hbox{\scriptsize max}}$ for ${\cal S}^3$
as in figure \ref{Fig:Sigma_distribution}, but display for ${\cal D}$ the
probability distribution of $\Sigma(\rho)$ with $\rho=36^\circ$.
It is seen that the distribution for ${\cal D}$ has been shifted to
smaller values of $\Sigma$, and thus the overlap of the distributions
for ${\cal D}$ and ${\cal S}^3$ has increased.

Analysing a sky map which possesses no pairs of circles,
one would obtain for randomly chosen circle pairs a distribution
corresponding to the shaded histograms in figures
\ref{Fig:Sigma_distribution} and \ref{Fig:Sigma_distribution_no_rot}.
Now the aim is to find the six pairs of circles due to the
dodecahedral topology.
If there would be only the pure ordinary SW contribution,
a value of $\Sigma_{\hbox{\scriptsize max}}=1$ would clearly stand out
of the distribution due to ${\cal S}^3$.
However, the Doppler and ISW contributions degrade this signal to
the grey histograms in figures \ref{Fig:Sigma_distribution}
and \ref{Fig:Sigma_distribution_no_rot}.
Thus the six pairs would probably yield values of
$\Sigma_{\hbox{\scriptsize max}}$ between 0.4 and 0.8
for $\Omega_{\hbox{\scriptsize tot}}=1.011$ and
between 0.4 and 0.7 for $\Omega_{\hbox{\scriptsize tot}}=1.020$,
and values of $\Sigma(\rho)$ between 0.3 and 0.75
for both values of $\Omega_{\hbox{\scriptsize tot}}$, respectively.
In the case of an analysis of a real sky map,
one would obtain many thousands of $\Sigma_{\hbox{\scriptsize max}}$ values
which would not correspond to  matched circles,
i.\,e.\ having a probability distribution looking like that of ${\cal S}^3$.
Embedded in this distribution, one would for the dodecahedron obtain
six $\Sigma_{\hbox{\scriptsize max}}$ values corresponding
to real matched circles.
Because of the large overlap of both distributions,
the signal of the six pairs of circles could easily be swamped.
Thus there is a non-vanishing probability
that the six pairs of matched circles can be overlooked in the sky maps.
The search for matched circles is in addition made more difficult by
the finite thickness of the SLS and
the Sunyaev-Zeldovich effect,
a possible systematic in the removal of foregrounds
and further secondary effects
which are not included in our analysis.

\section{Discussion and Summary}

Taking the WMAP value
$\Omega_{\hbox{\scriptsize tot}}=1.02\pm0.02$ as its face value,
we conclude that the Universe is slightly positively curved and
therefore necessarily spatially finite,
since all spherical space forms possess this property.
At present, the only way to decide on the particular topology
realized in our Universe consists in studying a concrete space form and
then compare its predictions with the observational data.

In this paper, we produce evidence in support of the hypothesis
that the spatial structure of the Universe could be given by the
Poincar\'e dodecahedral space ${\cal D}$,
provided that the total energy density parameter
$\Omega_{\hbox{\scriptsize tot}}$ is in the range $1.016\dots 1.020$.
For the detailed comparison with the WMAP data,
it was crucial that we could base our predictions for the CMB anisotropy
upon a large number of vibrational modes of the Poincar\'e dodecahedron.
Since the eigenfunctions of the Poincar\'e dodecahedron are not known
analytically, they have to be computed numerically.
Using the very efficient algorithm described in
section \ref{Numerical_determination}, we could take in our calculations
the first 10\,521 eigenfunctions into account.
In addition, we put forward the Conjecture (\ref{Eq:Conjecture})
for the homogeneous spherical space forms.
Combining the numerically computed eigenfunctions with the Conjecture,
we were able to calculate the CMB multipoles in a broad range,
see figures \ref{Fig:C_l} and \ref{Fig:delta_T_NSW_ISW},
covering the region of the large scale fluctuations.
Fixing the cosmological parameters as
$\Omega_{\hbox{\scriptsize bar}} = 0.046$ and
$\Omega_{\hbox{\scriptsize mat}} = 0.28$, $h=70$,
in agreement with the WMAP data, there are only two parameters left:
the amplitude $\alpha$ of the scale-invariant Harrison-Zeld'ovich spectrum
(\ref{Eq:Harrison_Zeldovich}), and the density parameter
$\Omega_{\hbox{\scriptsize tot}}>1$ or, equivalently, the curvature radius
$R_0 = \frac c{H_0}(\Omega_{\hbox{\scriptsize tot}}-1)^{-1/2}$.
(The dark energy is identified with a cosmological constant with magnitude
$\Omega_\Lambda := \Omega_{\hbox{\scriptsize tot}} -
\Omega_{\hbox{\scriptsize rad}} - \Omega_{\hbox{\scriptsize mat}}$.)
The parameter $\alpha$ was obtained from a fit to the $C_l$ values
as determined by WMAP in the range $l\in[20,45]$.
(This differs from the normalization in
\cite{Luminet_Weeks_Riazuelo_Lehoucq_Uzan_2003},
where $\alpha$ has been set to match the WMAP multipole with $l=4$ exactly,
since higher multipoles could not be calculated in
\cite{Luminet_Weeks_Riazuelo_Lehoucq_Uzan_2003} due to the very limited
number of available eigenfunctions.)
Having fixed in this way all parameters except
$\Omega_{\hbox{\scriptsize tot}}$,
we studied the CMB multipoles and the temperature correlation function
$C(\vartheta)$ for the dodecahedron as a function of
$\Omega_{\hbox{\scriptsize tot}}$.
One should keep in mind that we restricted the parameter space to the
above values.
Thus we cannot exclude that other choices could lead to different values
for $\Omega_{\hbox{\scriptsize tot}}$ due to geometrical degeneracies.

As a quantitative measure of the quality of our theoretical predictions,
we calculated the $S$ statistic (\ref{Eq:S-Statistik})
for the angles $\rho=60^\circ$ and $\rho=20^\circ$.
The result is given as a function of $\Omega_{\hbox{\scriptsize tot}}$
in figure \ref{Fig:S_statistic_median}
and shows for this statistic the remarkable feature of
pronounced minima at almost the same $\Omega_{\hbox{\scriptsize tot}}$ values,
$\Omega_{\hbox{\scriptsize tot}}\in[1.016,1.020]$.
In particular, figure \ref{Fig:S_statistic_median} shows
that the Poincar\'e dodecahedral universe leads for
$\Omega_{\hbox{\scriptsize tot}}$ in the above range to a suppression
of the CMB anisotropy at large scales in agreement with the WMAP data.
This is also demonstrated in figure \ref{Fig:C_theta}c),
which shows the temperature correlation function $C(\vartheta)$,
and in figures \ref{Fig:C_l}c) and \ref{Fig:delta_T_NSW_ISW}b)
showing the CMB power spectrum
($\Omega_{\hbox{\scriptsize tot}}=1.019$ is used in these figures).
Figure \ref{Fig:C_theta}c) demonstrates clearly that the dodecahedron
model describes the WMAP data much better than the concordance model.
At this point it is important to notice
that the suppression of the CMB anisotropy at large scales
occurring in the dodecahedron is neither obtained by fitting the
theoretical curves to one of the first (suppressed) multipoles
nor to the correlation function $C(\vartheta)$.
Rather the suppression is due to the finiteness of the Universe in
combination with a small volume of the dodecahedron,
$V({\cal D}) = \frac{\pi^2}{60} = 0.16449\dots$,
which is nicely illustrated in figure \ref{Fig:delta_T_NSW_ISW},
where we compare the power spectra of the dodecahedron ${\cal D}$
and the simply connected ${\cal S}^3$, respectively.
The structures seen in figures \ref{Fig:C_l} and \ref{Fig:delta_T_NSW_ISW}b)
in the CMB power spectrum at multipoles $l\lesssim 20$,
are mainly caused by the ordinary Sachs-Wolfe (SW) contribution
(dotted curve in \ref{Fig:delta_T_NSW_ISW}b))
whose analytic expression is given in eq.\,(\ref{Eq:multipol_NSW}).
The observed structures are a direct consequence of the
discrete mode spectrum of ${\cal D}$ possessing large gaps at
small wave numbers $\beta$,
see eqs.(\ref{Eq:E_beta}) and (\ref{Eq:Ikeda}).

The particularly striking suppression of the quadrupole and octopole
is also exemplified by figure \ref{Fig:C_23}
which displays the probability distribution $P(\delta T_l^2)$ ($l=2,3$)
obtained from sky simulations.
The probability distributions for the dodecahedron show
pronounced peaks which overlap nicely with the $1\sigma$ band
of the WMAP data.
In contrast, the concordance model predicts much broader distributions
producing large values for $\delta T_2^2$ and $\delta T_3^2$,
seen also in figure \ref{Fig:C_l}.

Summarising the results discussed so far,
it is fair to say that the Poincar\'e dodecahedral universe
provides a natural explanation  for the strange suppression of
power at large scales as observed by COBE and WMAP,
if $\Omega_{\hbox{\scriptsize tot}}$ is in the range $1.016\dots 1.020$,
which solves ``the mystery of missing fluctuations''.
At this point we would like to emphasise
that the good agreement of the Poincar\'e dodecahedral universe
with the WMAP data does by no means exclude that there are
other manifolds which lead to similar conclusions.
Rather, it may well be that there exists a ``topological degeneracy''
in addition to the well-known ``geometrical degeneracy''
(see e.\,g.\ \cite{Aurich_Steiner_2002b,Aurich_Steiner_2003}).
Since the value $\Omega_{\hbox{\scriptsize tot}}=1.02\pm0.02$
reported by the WMAP team depends on several priors and
includes the $1\sigma$ deviation uncertainty only,
there still exists the possibility of the Universe having
negative curvature.
In fact, e.\,g.\ the hyperbolic Picard space
\cite{Aurich_Lustig_Steiner_Then_2004a,Aurich_Lustig_Steiner_Then_2004b}
leads also to a suppression of the CMB fluctuations
at large scales in agreement with the WMAP data.

There remains to detect a more direct signature of the dodecahedral topology.
For this purpose, we have studied the circles-in-the-sky signature
in section \ref{circles-in-the-sky-signature}.
The dodecahedral universe possesses exactly 6 matched circles
with radii $42.7^\circ$ and $49.8^\circ$ for
$\Omega_{\hbox{\scriptsize tot}}=1.016$ and
$\Omega_{\hbox{\scriptsize tot}}=1.020$, respectively.
If there would be only the ordinary Sachs-Wolfe contribution,
one would predict an almost perfect matching of the circles.
However, as it is seen in figure \ref{Fig:delta_T_NSW_ISW},
there are important contributions from the integrated Sachs-Wolfe
and the Doppler effect.
The three contributions together destroy a perfect matching
as is illustrated for a given circle pair in figure
\ref{Fig:delta_T_NSW_ISW_phi}a).
While the ordinary Sachs-Wolfe contribution displayed in figure
\ref{Fig:delta_T_NSW_ISW_phi}b) shows the expected matching,
the Doppler and the integrated Sachs-Wolfe contributions shown
in figures \ref{Fig:delta_T_NSW_ISW_phi}c) and
\ref{Fig:delta_T_NSW_ISW_phi}d), respectively, differ appreciably
over the full angular range, and thus lead to the poor matching
demonstrated in figure \ref{Fig:delta_T_NSW_ISW_phi}a).

The quantity $\Sigma(\rho)$, eq.\,(\ref{Eq:S_measure}),
has been introduced for a search of matched circles in the
microwave sky.
Figures \ref{Fig:Sigma_distribution} and \ref{Fig:Sigma_distribution_no_rot}
display the probability distributions of $\Sigma_{\hbox{\scriptsize max}}$
and $\Sigma(36^\circ)$, respectively.
One observes broad distributions for both the dodecahedron ${\cal D}$
and the covering space ${\cal S}^3$.
While the distribution for ${\cal D}$ is peaked at higher values
of $\Sigma_{\hbox{\scriptsize max}}$ than the distribution for
${\cal S}^3$, the two distributions have a large overlap.
Due to the contamination by the Doppler and integrated Sachs-Wolfe effect,
the distribution for ${\cal D}$ has no peak at
$\Sigma_{\hbox{\scriptsize max}}\simeq 1$,
but rather gives for the most probable $\Sigma_{\hbox{\scriptsize max}}$
values at which the 6 matching circles are expected to be observed 
the interval $0.4\dots 0.8$.

In view of these complications, it seems that a search for matching circles
in the first-year WMAP data is quite difficult,
and thus the negative result reported in \cite{Cornish_Spergel_Starkman_1998b}
does not yet allow to conclude that the dodecahedral model is excluded.
We hope that the second-year WMAP data and the CMB measurements from
the Planck satellite will improve the situation considerably
such that we shall be able to decide about the topology of the Universe
in the near future.

Let us close with a quotation from Einstein,
who wrote in 1919 in describing the Universe of beings in a
``finite and yet unbounded Universe''
\cite{Einstein_1919,Einstein_2004}:
``The great charm resulting from this consideration lies in recognition of
the fact that {\it the universe of these beings is finite
and yet has no limits}.''


\section*{References}

\bibliography{../bib_chaos,../bib_astro}
\bibliographystyle{h-physrev3}

\end{document}